\documentclass[a4paper]{article}
\usepackage{jheppub}
\usepackage[utf8]{inputenc}
\usepackage[T1]{fontenc}

\usepackage{amsmath,amssymb,mathrsfs}
\usepackage{fancyhdr}
\usepackage{psfrag}
\usepackage{array,bbm}
\usepackage{url}
\usepackage{multirow}
\usepackage{dsfont}
\usepackage{cancel}

\usepackage{epsfig}

\numberwithin{equation}{section}

\newenvironment{Eqnarray}%
     {\arraycolsep 0.14em\begin{eqnarray}}{\end{eqnarray}}
\newcommand{\ba}{\begin{Eqnarray}}
\newcommand{\ea}{\end{Eqnarray}}
\newcommand{\be}{\begin{equation}}
\newcommand{\ee}{\end{equation}}

\def\half{\tfrac{1}{2}}
\def\centeron#1#2{{\setbox0=\hbox{#1}\setbox1=\hbox{#2}\ifdim
\wd1>\wd0\kern.5\wd1\kern-.5\wd0\fi
\copy0\kern-.5\wd0\kern-.5\wd1\copy1\ifdim\wd0>\wd1
\kern.5\wd0\kern-.5\wd1\fi}}
\def\ltap{\;\centeron{\raise.35ex\hbox{$<$}}{\lower.65ex\hbox{$\sim$}}\;}
\def\gtap{\;\centeron{\raise.35ex\hbox{$>$}}{\lower.65ex\hbox{$\sim$}}\;}

\def\lt{\left}
\def\rt{\right}

\title{One-loop contributions to neutral minima in the inert doublet model}
\author[a,b]{P.M.~Ferreira,}
\author[c]{Bogumi{\l}a~{\'S}wie{\.z}ewska}

\affiliation[a]{Instituto Superior de Engenharia de Lisboa, Portugal}
\affiliation[b]{Centro de F\'{\i}sica Te\'{o}rica e Computacional,
    Universidade de Lisboa, Portugal}
\affiliation[c]{Faculty of Physics, University of Warsaw, Pasteura 5, 02--093 Warsaw, Poland}

\emailAdd{ferreira@cii.fc.ul.pt}
 \emailAdd{bogumila.swiezewska@fuw.edu.pl}

\abstract{
The vacuum structure of the inert doublet model is analysed at the one-loop level
using the effective potential formalism, to verify the
validity of tree-level predictions for the properties of the global minimum. An inert minimum (with
massive fermions) and an inert-like minimum (with massless fermions) can coexist at tree level. But
the one-loop analysis reveals that the allowed parameter space for the coexistence of more than one
minimum is larger than the tree-level expected one. It is also shown that for some choices of
parameters, the global minimum found at the one-loop level may be inert (or inert-like), contrary
to what the tree-level analysis indicates.
}

\arxivnumber{1511.02879}

\begin{document}

\maketitle
\flushbottom

\section{Introduction}

The discovery of a Higgs boson~\cite{Atlas-Higgs, CMS-Higgs}, and the measurement of its
mass~\cite{Higgs-mass} have shed some light on the issue of vacuum stability, and consequently this
problem underwent detailed studies within the Standard Model
(SM)~\cite{Degrassi:2012, Buttazzo:2013, Spencer-Smith:2014, Kobakhidze:2014, Bednyakov:2015}.
However, it is well known that new interactions, which are necessary to solve some of the
problems faced by the SM, can  influence these considerations
significantly~\cite{Greenwood:2008, Branchina:2013prl, Branchina:2014sher, Lalak:2014}.

Among the possible extensions of the SM, and very interesting from the point of view of stability,
are models with an extended scalar sector. Additional scalars are commonly invoked to stabilise the
vacuum state --- since they contribute positively to the $\beta$ function of the running Higgs
self-coupling $\lambda$, they prevent $\lambda$ from turning negative (or at least postpone it to
high energies), see
e.g.~\cite{Nie:1998yn,Ferreira:2009jb,Kannike:2011, Goudelis:2013, Khan:2015, Chakrabarty:2015,
Ferreira:2015rha}, but they can also affect its behaviour at lower energies~\cite{Swiezewska:2015}.
On the other hand, with more scalar fields the vacuum structure of extended models becomes
nontrivial, even at tree level, since there exists more than one direction in scalar space along
which vacuum expectation values (vevs) may develop. Thus a possibility of spontaneous CP or charge violation arises.

Among the simplest scalar extensions of the SM are the two-Higgs-doublet models (2HDMs), for a
recent review see~\cite{2hdm-rev}. In 2HDMs there are two scalar doublets, which gives eight real scalar
fields. With this field content, many different vacuum states are possible, depending on which of
the fields develop a non-zero vev. The tree-level vacuum structure of 2HDMs has been well
studied~\cite{Ma:1978, Ferreira:2004, Barroso:2005, Ivanov:2006, Ivanov:2007, Barroso:2007,
Barroso:2013, Ivanov:2015, Ginzburg:2010}. In particular it has been shown that, already at tree
level, two different minima {\em of the same nature, which break the same symmetries}, can coexist,
e.g. two normal minima (i.e. two minima at which the neutral components of the doublets have
non-zero vevs). This brings about the possibility of a metastable vacuum state.

In the present work we focus on the inert doublet model (IDM), which is a special realisation of
2HDM that exactly preserves a $\mathbb{Z}_2$ symmetry, before and after spontaneous symmetry
breaking. The model was first introduced by Ma in 1978~\cite{Ma:1978}, and then rediscovered in
2006 as a possible way of improving the naturalness of the SM by incorporating a heavy
Higgs boson~\cite{Barbieri:2006}, and as a source of dark matter (DM)~\cite{Cao:2007}. The DM
candidate of the IDM has then been shown to account for the observed relic density of
DM~\cite{LopezHonorez:2006,Dolle:2009, Honorez:2010,Sokolowska:2011,Gustafsson:2012}, and to be
in agreement with direct detection limits~\cite{Arhrib:2013,Klasen:2013, Abe:2014, Abe:2015, Ilnicka:2015}.
Moreover, the model is particularly interesting since the Higgs boson of the IDM is SM-like,
which is in agreement with current LHC data~\cite{Swiezewska:2012,Krawczyk:2013jhep, Arhrib:2013,
Goudelis:2013,Chakrabarty:2015, Kraml:2015, Ilnicka:2015}. The IDM has been also studied in the context of the
thermal evolution of the Universe~\cite{Ginzburg:2010, Sokolowska:2011, Sokolowska:2011t}, and in
particular baryogenesis, as a strong first order phase transition can occur within this
model~\cite{Kanemura:2004, Gil:2012, Cline:2013,Chowdhury:2011, Blinov:2015}. Extending the IDM
slightly, one can also account for the observed neutrino
masses~\cite{Ma:2006neutrino, Gustafsson:2012neutrino,Chakrabarty:2015, Merle:2015, Longas:2015}.

The vacuum state of the IDM (the inert vacuum) is realised when only one of the doublets acquires a
non-zero vev. In this case, there is still the possibility of a coexisting inert-like minimum, with
the other doublet having a non-vanishing vev~\cite{Ginzburg:2010}. In order to preserve the
$\mathbb{Z}_2$ symmetry which stabilises the DM
 candidates, in the IDM only the SM-like
doublet couples to fermions. Therefore, even though both minima yield similar physics in the scalar
and gauge sectors, their behaviour is markedly different when one takes into account fermions --- the
inert vacuum gives fermions their masses through the usual Higgs mechanism, but in the inert-like
vacuum all fermions are necessarily massless. Thus the inert-like vacuum is unphysical and should be
avoided. At tree level it is easy to compare the depths of both minima, and one
can even establish analytical formulae for the difference of the values of the potential at both
minima. Those tree-level relations relating the value of the potential at the inert and inert-like
vacua can be written in several (equivalent at tree level) ways, in particular in such a form that
they depend only on the masses of physical particles in the two minima, and on the two vevs. Then,
by adjusting the values of the parameters, we can ensure that the inert minimum is the global one,
thus avoiding the situation where it would be metastable.

The question we wish to address in the present paper is as follows: what is the impact of the loop
corrections on the relation between inert and inert-like minima? Could the loop corrections change
the depth of the two minima in such a way that the minimum that was the global one at
tree level, turns to a local one at one-loop level? The answer to this question is crucial for the stability of
the tree-level considerations of the IDM against loop corrections. Further, we will investigate in detail
the validity, at one-loop, of the formulae established at tree level for the relative depths of the potentials.

The tool to study vacua beyond tree level is the one-loop effective potential. In contrast to the
tree-level scalar  potential, it receives contributions from all the particles present in the model: scalars,
gauge bosons, and fermions. It will be instructive to see what is the impact of each sector on the issue of the
stability of the inert vacuum, and thus in our study we incorporate the respective sectors sequentially, so that
the influence of each of them can be clearly seen. In order to establish whether a solution of the one-loop
minimisation conditions is a minimum, we will require expressions for the one-loop squared scalar masses
in the IDM. We will therefore compute the scalars' one-loop self energies and present analytical formulae
for the one-loop masses.

The one-loop effective potential of the IDM was thoroughly analysed in
ref.~\cite{Gil:2012}. The main purpose of that work was to study the possibility of a first-order phase
transition in the course of the evolution of the Universe, for which a temperature-dependent effective
potential was used. Nonetheless, the authors of~\cite{Gil:2012} also studied the zero-temperature potential
which we will analyse in the current paper. Our particular emphasis is in the relationship between vacua,
which they did not directly discuss, but we will see that the conclusions we derive are
in agreement with their findings.

We will discover that, though the tree-level formulae relating the depth of the minima are a very good
approximation to the one-loop results, there are still considerable differences to be found, specially when
one takes into account the fermions of the theory. Specifically, we will discover that for some regions of
parameter space the tree-level predictions for the nature of the global minimum can be contradicted by the
one-loop results. We will also show that with the use of the one-loop potential the parameter space for which inert
and inert-like minima can coexist is enlarged.

This paper is organised as follows: in section~\ref{sec:inv} we review the 2HDM scalar potential which gives
rise to the IDM. We pay special attention to its scalar potential, reviewing the conditions under which
inert and inert-like vacua can coexist, as well as the vacuum stability bounds one must impose on the potential.
The tree-level relationships between the depths of the potential at different vacua are reviewed, as well as
the Yukawa interactions of the model, which gives different fermionic physics in both minima. In
section~\ref{sec:onel} the one-loop effective potential which we will be using is introduced, and the
computation of the one-loop scalar self energies and masses is undertaken for the inert vacuum. We generalise
these results for the inert-like vacuum as well. In section~\ref{sec:comp} we then compare the one-loop
inert and inert-like minima, performing a vast scan of the IDM's parameter space and investigating the validity
of the tree-level formulae which relate the depths of the potential at different minima. We analyse those
results in section~\ref{sec:and} and draw conclusions in section~\ref{sec:conc}.

\section{The inert doublet model at tree level\label{sec:tree-level}}
\label{sec:inv}

The 2HDM scalar potential for inert models is given by
\ba
 V_0 &=& m^2_{11}\Phi_1^\dagger\Phi_1+m^2_{22}\Phi_2^\dagger\Phi_2
      +\half\lambda_1\,|\Phi_1|^4
   +\half\lambda_2\,|\Phi_2|^4 +\lambda_3\,|\Phi_1|^2\,|\Phi_2|^2  \\
   && +\lambda_4\,|\Phi_1^\dagger\Phi_2|^2
   +\half\lambda_5\lt[\lt(\Phi_1^\dagger\Phi_2\rt)^2+{\rm h.c.}\rt]\,,
\label{eq:pot}
\ea
where all parameters are real (including $\lambda_5$, see below). The potential is $\mathbb{Z}_2$-symmetric under a transformation that changes the sign of either $\Phi_1$ or $\Phi_2$.

Unless specific conditions are imposed on the quartic couplings, there may exist
directions in field space along which the value of the potential tends to minus infinity --- a concern which already
exists in the SM. We must therefore ensure that the scalar potential is bounded from below --- thus guaranteeing the
 existence of at least a stable minimum --- and this is accomplished if the quartic
couplings obey the following conditions:
\be
\lambda_1 > 0\;, \quad  \lambda_2 > 0\;, \quad
\lambda_3 > -\sqrt{\lambda_1 \lambda_2}\;, \quad
\lambda_3 + \lambda_4 - |\lambda_5| > -\sqrt{\lambda_1 \lambda_2}\;.
\label{eq:bfb}
\ee
It was shown in~\cite{Ivanov:2006,Ivanov:2007} that these are necessary and sufficient conditions to guarantee
that the potential is bounded from below. 
These are tree-level conditions, so in principle they do not have to be valid at one-loop level. However, in the perturbativity range of the theory the one-loop corrections to the effective potential are small and should not affect significantly the asymptotic behaviour of the potential. Since we will be working at a fixed value of the renormalisation scale (see below), for which the loop corrections are small, using the tree-level constraints at this scale is consistent  with the one-loop analysis (see also~\cite{Goudelis:2013,Sher:1988,Nie:1998yn,Ferreira:2009jb,Ferreira:2015rha}).

The Yukawa sector has to be specified, and the $\mathbb{Z}_2$ symmetry must be applied to the whole lagrangian.
In fact, the application of the $\mathbb{Z}_2$ symmetry to the Yukawa sector of the 2HDM to prevent the occurrence of
tree-level Higgs-mediated flavour changing neutral currents (FCNC) was the reason this symmetry was first
proposed by Glashow, Weinberg and Paschos~\cite{Glashow:1976nt,Paschos:1976ay}. FCNC are avoided if
only one of the doublets couples to fermions of the same charge, which is easily accomplished if one
chooses the $\mathbb{Z}_2$ ``charges''
of the fermions appropriately.
For the $\mathbb{Z}_2$-symmetric 2HDM, here are several possibilities explored in the
literature: in model type I,
only $\Phi_2$ couples to all fermions; in model type II, $\Phi_2$ couples to up-type quarks and $\Phi_1$ to
the remaining fermions; in model type X, all quarks couple to $\Phi_1$ but the leptons couple to $\Phi_2$; and in
model type Y, up-type quarks and leptons couple to $\Phi_2$, while down-type quarks couple to $\Phi_1$.
In the IDM, traditionally, the doublet which is made to couple to all fermions is $\Phi_1$, and
$\Phi_2$ has no coupling to fermions at all.\footnote{The choice of $\Phi_1$ or $\Phi_2$ is a matter of
convention, there is no physical distinction, before this choice, between the doublets.} Therefore, keeping
only the third-generation fermions, the Yukawa lagrangian for the IDM is given by
\be
-\,{\cal L}_Y \,=\,\lambda_t \,\bar{Q}_L\,\tilde{\Phi}_1\,t_R\,+\, \lambda_b \,\bar{Q}_L\,\Phi_1\,b_R\,+\,
\lambda_\tau \,\bar{L}_L\,\Phi_1\,\tau_R\, ,
\label{eq:lyuk}
\ee
where $Q_L$ and $L_L$ are $SU(2)_L$ doublets for fermions and leptons, respectively, $t_R$, $b_R$ and $\tau_R$
are the right-handed fields for the top, bottom and tau fermions, the $\lambda_i$ the respective Yukawa
couplings
and $\tilde{\Phi}_1$ is the charge-conjugate of the doublet $\Phi_1$.\footnote{We are assuming
zero masses for neutrinos.} One of the consequences of having a single doublet coupling to fermions is
that all of the parameters
in the scalar potential~\eqref{eq:pot} can be made real through an appropriate rephasing of the fields
(this would happen even if the Yukawa sector was similar to 2HDM's type II, X or Y, as well).
 Note however, that by choosing this Yukawa lagrangian we single out the $\Phi_1$ doublet as the one
 coupling to fermions, and therefore the full lagrangian is $\mathbb{Z}_2$-symmetric under the exchange of the sign of $\Phi_2$, and not $\Phi_1$.

For the model to be complete, a vacuum state is needed. Since the potential of eq.~\eqref{eq:pot} is bounded from
below, it will certainly have a minimum for some value of the fields.
The scalar doublets $\Phi_i$ contain eight real scalar fields, which we parameterise as
\be
\Phi_1 = \frac{1}{\sqrt{2}}\,\left(\begin{array}{c} c_1 + \mathrm{i} \,c_2 \\ r_1 +
\mathrm{i} \,i_1\end{array} \right)\;\;\;
,\;\;\; \Phi_2 = \frac{1}{\sqrt{2}}\,\left(\begin{array}{c} c_3 + \mathrm{i} \,c_4 \\
r_2 + \mathrm{i}\, i_2\end{array} \right)
\label{eq:doub}
\ee
%
($\mathrm{i}$ is the imaginary unit and $i_j$ the neutral, imaginary field components).
We are interested in electroweak-symmetry-breaking minima, so we exclude the trivial extrema which
form at the origin (with the vevs of all component fields equal to zero). The 2HDM can have minima which
break charge conservation, if the doublets acquire vacuum
expectation values on their upper components. In this paper we will only study neutral minima. The 2HDM
can equally develop CP-breaking minima, if the doublets have complex vevs --- but it has been
shown~\cite{Branco:1985aq,Ferreira:2010hy}
that for a 2HDM with an exact $\mathbb{Z}_2$ symmetry such vacua cannot occur. We are thus left with
minima where the doublets
have real vevs in their neutral components, such that
$\langle\Phi_1 \rangle = ( 0 \, , \, v_1)^T/\sqrt{2}$ and
$\langle\Phi_2 \rangle = ( 0 \, , \, v_2)^T/\sqrt{2}$. There are still, however, three possibilities
for extrema, leading to different physics, depending on the values of the potential's parameters:

\begin{description}
\item[``Normal'' 2HDM extremum.] In the ``normal'' extremum both $v_1$ and $v_2$ are non-zero
and therefore the vacuum breaks the $\mathbb{Z}_2$ symmetry of the potential.
This extremum has been the target of intensive studies (for instance, ~\cite{Ma:1978,Ferreira:2004, Barroso:2005,
Ivanov:2006, Ivanov:2007, Barroso:2007,Barroso:2013, Ivanov:2015, Ginzburg:2010, Ferreira:2015rha,
Barroso:2012}, and many references in~\cite{2hdm-rev}), however we will not study it in this work.
\item[Inert extremum.] In the inert extremum the vev $v_1$ is non-zero and $v_2 = 0$. Therefore, with
an inert vacuum state the $\mathbb{Z}_2$ symmetry of the potential and the Yukawa terms is preserved after spontaneous symmetry breaking.
This is the state
selected as the vacuum state of the IDM.

The real neutral component of $\Phi_1$ corresponds to the Higgs scalar $h$
discovered at the LHC --- its couplings to fermions and gauge bosons are indeed identical to those of
the SM Higgs. The remaining components of $\Phi_1$ originate the Goldstone bosons $G$ and $G^\pm$,
and the component fields of the doublet $\Phi_2$  give rise to other scalars --- 
neutral $H$ and $A$ and a charged scalar $H^\pm$. None of the extra scalars couple
 to fermions, and the lightest of them is a
dark matter candidate --- the intact $\mathbb{Z}_2$ symmetry originates a conserved quantum number,
which implies that the scalars
within the second doublet are always produced in pairs. Not coupling to fermions either, they are perfect
dark matter candidates, also called ``inert scalars''. Usually dark matter studies
within the IDM consider only the regions of parameter space for which the lightest inert scalar is
neutral.

Since the inert vacuum is characterised by $\langle r_1\rangle=v_1$
($v_1 = v$ = 246 GeV)  and all other fields having a
zero vev
at tree level, the scalars'
mass matrix is diagonal, and the scalar states have mass eigenvalues
 (before applying the minimisation conditions)
given by
\ba
m^2_{G_0} \;=\; \left(\frac{\partial^2 V_0}{\partial i_1^2}\right)_{I}
&=& m_{11}^2 + \frac{1}{2}\,\lambda_1\,v_1^2,
\label{eq:mG0} \\
m^2_{G^\pm_0} \;=\; \left(\frac{\partial^2 V_0}{\partial c_1^2}\right)_{I} = \left(\frac{\partial^2 V_0}{\partial c_2^2}\right)_{I}
&=& m_{11}^2 + \frac{1}{2}\,\lambda_1\,v_1^2,
\label{eq:mGch0} \\
m^2_{h_0} \;=\; \left(\frac{\partial^2 V_0}{\partial r_1^2}\right)_{I}
&=& m_{11}^2 + \frac{3}{2}\,\lambda_1\,v_1^2,
\label{eq:mh0} \\
m^2_{H_0} \;=\; \left(\frac{\partial^2 V_0}{\partial r_2^2}\right)_{I}
&=& m_{22}^2 + \frac{1}{2}\,\lambda_{345}\,v_1^2,
\label{eq:mH0}\\
m^2_{A_0} \;=\; \left(\frac{\partial^2 V_0}{\partial i_2^2}\right)_{I}
&=& m_{22}^2 + \frac{1}{2}\,\bar{\lambda}_{345}\,v_1^2,
\label{eq:mA0} \\
m^2_{H^\pm_0} \;=\; \left(\frac{\partial^2 V_0}{\partial c_3^2}\right)_{I} = \left(\frac{\partial^2 V_0}{\partial c_4^2}\right)_{I}
&=& m_{22}^2 + \frac{1}{2}\,\lambda_3 \,v_1^2,
\label{eq:mHch0}
\ea
with the notation $\lambda_{345} = \lambda_3 + \lambda_4 +\lambda_5$ and
$\bar{\lambda}_{345} = \lambda_3 + \lambda_4 - \lambda_5$.
 As for the fermions
(third generation only) and gauge bosons, at tree level their masses are given by
\begin{align}
m_{t_0} &= \frac{\lambda_t}{\sqrt{2}}\,v_1\, , & m_{b_0} &= \frac{\lambda_b}{\sqrt{2}}\,v_1\, , &
m_{\tau_0} &= \frac{\lambda_\tau}{\sqrt{2}}\,v_1\, , \label{eq:mf0} \\
m^2_{W_0} &= \frac{g^2}{4}\,v_1^2 \, , & m^2_{Z_0} &= \frac{g^2 + {g^\prime}^2}{4}\,v_1^2 \, , &
\label{eq:mfG}
\end{align}
with the usual SM Yukawa couplings $\lambda_i$ and gauge couplings $g$ and $g^\prime$.

The potential~\eqref{eq:pot} has 7 independent real
parameters. One usually ``trades'' the potential parameters, whenever possible, for potentially observable
quantities --- meaning, the four physical masses ($m_h$, $m_H$, $m_A$ and $m_{H^\pm}$, bearing in mind that we
already know from LHC that $m_h \simeq 125$ GeV) and the vev $v = 246$ GeV. This way we are left with two
undetermined parameters, which we will take to be $\lambda_2$ and $m_{22}^2$.

\item[Inert-like extremum.] In the inert-like extremum the vev $v_2$ is non-zero and $v_1 = 0$.
The $\mathbb{Z}_2$
symmetry of the potential is preserved by this state, however the
$\mathbb{Z}_2$ symmetry of the lagrangian
is spontaneously
violated.

 The masses of the scalar particles and the gauge bosons are given by the
 formulas~\eqref{eq:mG0}--\eqref{eq:mHch0}, and \eqref{eq:mfG}, with the exchange $1\leftrightarrow2$
 in scalar couplings' and vevs' indices.\footnote{One should note, however, that the particles called the same names in the inert and inert-like minima, in fact follow from different doublets. So in the inert-like minimum $m_{G_0}^2=\left(\frac{\partial^2 V_0}{\partial i_2^2}\right)_{IL}$, $m_{h_0}^2=\left(\frac{\partial^2 V_0}{\partial r_2^2}\right)_{IL}$, etc.}
The fermions are massless in such a minimum, since only the doublet
$\Phi_1$ couples to fermions. As such this vacuum is not physical and does not describe our Universe,
and the choices of parameters for which this vacuum is the global minimum of the model should be avoided.
\end{description}

 Notice that the Yukawa lagrangian, eq.~\eqref{eq:lyuk}, is a feature of the model,
 {\em regardless of the vacuum the theory
finds itself in}. In other words, {\em the Yukawa couplings are the same in both inert and inert-like vacua}, even though their values are determined by the fermion masses in the inert minimum. This will be important later on, when we compute the one-loop masses at the inert-like vacuum.

Minimising the scalar potential~\eqref{eq:pot}, one finds that for stationary points such as the inert one to
occur, one has
\be
v_1^2 \,=\,-\,\frac{2\,m_{11}^2}{\lambda_1},\;\;\;\mbox{provided}\;\;\;m_{11}^2<0.
\ee
The definite sign imposed on the $m^2_{11}$ parameter is of course a consequence of the bounded from below
condition on $\lambda_1$, from eq.~\eqref{eq:bfb}.
These conditions do not guarantee a minimum, one would have to verify that all squared scalar masses
are positive, which would impose further constraints on the model's parameters.
Similarly, an inert-like stationary point may occur if one has
\be
v_2^2 \,=\,-\,\frac{2\,m_{22}^2}{\lambda_2},\;\;\;\mbox{provided}\;\;\;m_{22}^2<0.
\ee

In the 2HDM at tree level, it has been proven that minima which break different symmetries of the potential cannot
coexist~\cite{Ivanov:2006,Ivanov:2007,Barroso:2007}.
What this means, for instance, is that ``normal'' minima cannot coexist with CP or charge breaking minima --- if
one such minimum exists, the other stationary points, if they occur, are guaranteed to be saddle
points. Similarly, if the potential's parameters are such that a minimum with non-zero $(v_1\,,\,v_2)$
(therefore breaking the $\mathbb{Z}_2$ symmetry of the potential) exists, then no inert or inert-like minimum exists.
However, both inert and inert-like extrema preserve the $\mathbb{Z}_2$ symmetry of the potential, and as such
can, in principle,
exist simultaneously. There are however conditions for such simultaneous minima, which are given by, at
tree level:
\be
 \mbox{{\em Inert and inert-like minima can coexist in the potential if}} \;\;m_{11}^2<0 \;\;\mbox{{\em
 and}}\;\; m_{22}^2<0.
\label{eq:sta}
\ee
Notice that this is only a {\em necessary} condition, not a {\em sufficient} one. For instance, another
necessary tree-level condition for simultaneous inert and inert-like minima is that $\lambda_3 + \lambda_4 +
\lambda_5 > 0$ (see eq.~\eqref{eq:mH0}).

We can now ask what is the relation between the depths of the potential at inert  ($V_I$)
and inert-like minima ($V_{IL}$). It is possible to find analytical expressions for the difference in
depths of the potential at these minima,  we have
\ba
V_I \,-\,V_{IL} &=& \frac{1}{2} \left( \frac{m^4_{22}}{\lambda_2} \,-\,
\frac{m^4_{11}}{\lambda_1}\right)
\label{eq:difV01} \\
 & & \nonumber \\
 & = & \frac{1}{4} \left[ \left( \frac{m^2_{H^\pm}}{{v_2}^2}\right)_{IL} \,-\,
   \left(\frac{m^2_{H^\pm}}{v_1^2}\right)_{I}\right]\,v_1^2\,{v_2}^2 \;\;\; .
\label{eq:difV02}
\ea
In the second formula we have used the values of the squared charged masses at each minimum. Notice that
these formulae do not privilege any of the two types of minima, inert or inert-like. Depending on the choice
of parameters for the potential, one or the other might be the global minimum of the theory.
 We stress that these
formulae were deduced at {\em tree level}, and their validity beyond that has not, up until now, been verified.
And that is the main objective of the current work: can quantum corrections to the potential affect
the relative depth of the potential at both minima? Is it possible that, for some choices of parameters, one
may have an ``inversion'' of minima --- the minimum predicted to be global at tree level turning out to be
a local one when loop corrections are considered? To study these questions we require the one-loop potential
and, for consistency, the one-loop masses for the scalars of the model.

\section{The one-loop 2HDM potential and scalar masses}
\label{sec:onel}

Since we are interested in investigating one-loop results, a consistent renormalisation
strategy must be followed. We adopted the procedure outlined by Martin
in~\cite{Martin:2003qz,Martin:2003it}. Briefly, it consists of solving the minimisation
equations obtained from the one-loop effective potential to determine the vevs of the theory,
and computing the one-loop self energies for all scalar particles. The loop calculation
uses a mass-independent renormalisation scheme with minimal subtraction. The input parameters of
this scheme are the renormalised scalar couplings of the theory, as well as the renormalised
running masses of fermions and gauge bosons. The outputs are therefore the physical observables,
such as the scalars' masses,
cross sections and so on.

The starting point to define the one-loop minimum is
therefore the effective potential, which is given by
\be
V\,=\,V_0 \,+\, V_1\,,
\label{eq:Vt}
\ee
with $V_0$ given by eq.~\eqref{eq:pot} and the one-loop contribution equal to
\be
V_1 = \frac{1}{64\pi^2}\, \sum_\alpha n_\alpha \,m^4_\alpha(\varphi_i) \left[\log\left(\frac{m^2_\alpha(\varphi_i)}{\mu^2}\right)\,-\,\frac{3}{2}\right],
\label{eq:V1}
\ee
where $\mu$ is the renormalisation scale chosen and the $m_\alpha(\varphi_i)$ are the
field-dependent tree-level mass eigenvalues of all particles present in the
theory.\footnote{They depend on the
eight real components of the doublets, which we represented by $\varphi_i$.} We have chosen
to work in the Landau gauge and use dimensional reduction (DRED) as our
regularisation method. The sum in $\alpha$ runs over all particles present in the theory, meaning:
all the scalar eigenstates, taking into account that some of them are degenerate, such as the two
charged Goldstones and charged scalars;
all the fermions, taking into account their spin, colour and charge degrees of freedom; and likewise
all gauge bosons.\footnote{At this stage, the most obvious difference in using DRED rather
than dimensional regularisation (DREG) is the common factor of $3/2$ present in all terms of the sum in
eq.~\eqref{eq:V1} --- in DREG that factor would be different for gauge bosons.}
Thus the factor $n_\alpha$ counts the number of degrees of freedom corresponding to each particle,
and it is given, for a particle of spin $s_\alpha$, by
\be
n_\alpha \,=\, (-1)^{2s_\alpha}\,Q_\alpha C_\alpha (2 s_\alpha + 1),
\ee
where $Q_\alpha$ is 1 for uncharged particles and 2 for charged ones; $C_\alpha$ counts the number of
colour degrees of freedom (for particles without colour it equals 1, for particles with colour, 3).
It is well known that the one-loop effective potential is a gauge-dependent entity, but its value
at any stationary point is a physical quantity~\cite{Nielsen:1975fs}. Since we are interested in studying
the depths of the potential in two different minima at one-loop, we can therefore trust the results
obtained from~\eqref{eq:Vt}.

To compute the one-loop potential we need expressions for the field-dependent tree-level masses of
all particles present in the theory. In the inert/inert-like minima they depend only on the field $r_1$/$r_2$
as was shown in section~\ref{sec:tree-level}, so the formulas for mass-eigenvalues are simple, and would be given by
the formulas~\eqref{eq:mG0}--\eqref{eq:mfG}, with $v_1$ exchanged to $r_1$ in the inert state, and analogous
changes made for the inert-like minimum (bearing in mind that the fermions will be massless). In practice,
for our purposes it is sufficient to simply write all masses in terms of the vevs $v_1$ and $v_2$.

\subsection{Inert vacua}

The first derivatives of the one-loop potential~\eqref{eq:Vt} specify the location of
the one-loop minimum, and they are given by (dropping the
explicit field dependence in the masses for simplicity of notation)
\be
\frac{\partial V}{\partial \varphi_i} = \frac{\partial V_0}{\partial \varphi_i} \,+\,
\frac{1}{32\pi^2}\, \sum_\alpha n_{\alpha}m^2_\alpha \,\frac{\partial m^2_\alpha}{\partial \varphi_i}
\left[\log\left(\frac{m^2_\alpha}{\mu^2}\right)\,-\,1\right].
\label{eq:dV}
\ee
Equation~\eqref{eq:dV} can be considerably simplified for any computation at the inert (and inert-like) extremum.
In the inert case, we have $\langle r_1\rangle = v_1$, and all remaining $\varphi_i = 0$. An explicit
calculation has shown that all derivatives of the one-loop potential with respect to all but one of the fields
$\varphi_i$ are then trivially equal to zero for this case --- the only non-trivial derivative is
$\partial V/\partial r_1$.
This, as in the tree-level case, is due to the quadratic plus quartic form of
the potential.
Due to the conventions we have chosen, performing this derivative is equivalent to
differentiating with respect to $v_1$. Taking advantage of the expressions~\eqref{eq:mG0}--\eqref{eq:mfG},
we obtain (in the Landau gauge and using DRED)
\ba
\frac{1}{v_1}\frac{\partial V}{\partial v_1} & = &  m_{11}^2 + \frac{1}{2}\,\lambda_1\,v_1^2 \,
\nonumber \\
 & +& \frac{1}{32\pi^2} \,\left\{
 \lambda_1\,m^2_{G_0}\,\left[\log\left(\frac{m^2_{G_0}}{\mu^2}\right)\,-\,1\right] \,+\,
 3\,\lambda_1\,m^2_{h_0}\,\left[\log\left(\frac{m^2_{h_0}}{\mu^2}\right)\,-\,1\right] \,+\right.
 \nonumber \\
 & +&  \lambda_{345}
 \,m^2_{H_0}\,\left[\log\left(\frac{m^2_{H_0}}{\mu^2}\right)\,-\,1\right] \,+\, \bar{\lambda}_{345}
 \,m^2_{A_0}\,\left[\log\left(\frac{m^2_{A_0}}{\mu^2}\right)\,-\,1\right] \,
 \nonumber \\
 & +&   2\,\lambda_3\,m^2_{H^\pm_0}\,\left[\log\left(\frac{m^2_{H^\pm_0}}{\mu^2}\right)\,-\,1\right] \,+\,
 2\,\lambda_1\,m^2_{G^\pm_0}\,\left[\log\left(\frac{m^2_{G^\pm_0}}{\mu^2}\right)\,-\,1\right] \,
 \nonumber \\
 & -& 6\,\lambda_t^2\,m^2_{t_0}\,\left[\log\left(\frac{m^2_{t_0}}{\mu^2}\right)\,-\,1\right] \,-\,
 6\,\lambda_b^2\,m^2_{b_0}\,\left[\log\left(\frac{m^2_{b_0}}{\mu^2}\right)\,-\,1\right] \,-\,
 2\,\lambda_\tau^2\,m^2_{\tau_0}\,\left[\log\left(\frac{m^2_{\tau_0}}{\mu^2}\right)\,-\,1\right] \,
  \nonumber \\
 & +& \left.
 3\,\frac{g^2 + {g^\prime}^2}{2}\,m^2_{Z_0}\left[\log\left(\frac{m^2_{Z_0}}{\mu^2}\right)\,-\,1\right] \,+\,
 3\,g^2\,m^2_{W_0}\left[\log\left(\frac{m^2_{W_0}}{\mu^2}\right)\,-\,1\right]
\right\} \,=\, 0.
\label{eq:dVt}
\ea
To verify the existence of an inert-like vacuum, one needs to write all particle masses in terms
of  $\langle r_2\rangle = v_2$,
and perform the derivative of the
potential with respect to $v_2$. The expression for that derivative may be obtained from eq.~\eqref{eq:dVt}
with the following prescriptions: set to zero all contributions from fermions (in the inert-like vacuum
they are massless) and perform the swap $1 \leftrightarrow 2$ in the indices of the several
scalar couplings and vevs.

The potential's first derivatives allow one to verify the existence of an inert or inert-like stationary
points. To verify that they are minima, however, one needs the one-loop potential's second derivatives,
{\em i.e.}
the masses of all scalars in the model.\footnote{In fact, the second derivatives of the potential are
the one-loop
masses computed at zero external momentum.} We have computed the one-loop self energies for the IDM
in two separate ways, as a cross check. We compared a direct diagrammatic calculation with an
adaptation of the results of Martin~\cite{Martin:2003qz,Martin:2003it}, which provide formulae for the
self energies of scalars in a theory with arbitrary number of scalars,
generic gauge couplings and Yukawa interactions. In order to check that our calculations are correct, we
verified that we obtain the correct number of Goldstone bosons ({\em i.e.} that a charged scalar and a neutral
one are massless at a stationary point of the model). We also verified that our solutions coincide
with those
obtained in the effective potential approximation for the neutral scalars.\footnote{The effective potential
approximation takes as the scalar masses the eigenvalues of the matrix of the second derivatives of the
one-loop potential with respect to the scalar fields. As was already mentioned, this in fact corresponds to
an approximation, the physical masses computed at zero external momentum. This is however a very good
approximation~\cite{Ellis:1990nz,Ellis:1991zd,Brignole:1991pq}, and an easy one to obtain, thus providing a
very good check of the exact calculations.}

The computation of the one-loop contributions to the mass of the scalars at an inert minimum gives
(adapting the results from Martin~\cite{Martin:2003qz,Martin:2003it}):
\subsection*{For the non-inert scalar $h$:}
\be
m^2_h \,=\,m^2_{h_0} \,+\,\frac{1}{32\pi^2}\,m^2_{h_1},
\label{eq:mh}
\ee
with $m^2_{h_0}$ given by eq.~\eqref{eq:mh0} and we have divided the one-loop contribution into three parts,
\be
m^2_{h_1} \;=\; \mbox{Re}(m^2_{h_{1,S}} \,+\, m^2_{h_{1,G}} \,+\, m^2_{h_{1,F}})\, ,
\ee
with $m^2_{h_{1,S}}$ collecting contributions from self-energy diagrams containing only internal scalar lines,
$m^2_{h_{1,G}}$ referring to diagrams with at least one internal gauge boson line and $m^2_{h_{1,F}}$
collecting the contributions from diagrams with internal fermion lines. We then have
\ba
m^2_{h_{1,S}} &=& \lambda_1 A(m_{G_0}) \,+\,2\lambda_1 A(m_{G^\pm_0}) \,+\,3\lambda_1 A(m_{h_0}) \,+\,
\lambda_{345} A(m_{H_0}) \,+\,\bar{\lambda}_{345} A(m_{A_0}) \,+\, 2\lambda_3 A(m_{H^\pm_0})
\vspace{0.4cm} \nonumber \\
 & + & \lambda_1^2 \,v_1^2 \,B(m_{G_0},m_{G_0},p^2) \,+\, 2 \lambda_1^2 \,v_1^2\,
 B(m_{G^\pm_0},m_{G^\pm_0},p^2) \,+\,
  9\lambda_1^2 \,v_1^2\, B(m_{h_0},m_{h_0},p^2)
  \vspace{0.4cm} \nonumber \\
  & + & \lambda_{345}^2 \,v_1^2\, B(m_{H_0},m_{H_0},p^2) \,+\,
  \bar{\lambda}_{345}^2 \,v_1^2\, B(m_{A_0},m_{A_0},p^2)
  \,+\, 2\lambda_3^2 \,v_1^2 \,B(m_{H^\pm_0},m_{H^\pm_0},p^2) \,,
  \label{eq:h1}
\ea
where we use the Passarino--Veltman functions~\cite{Passarino:1978} in the
$\overline{\textrm{MS}}$/$\overline{\textrm{DR}}$ scheme,
\be
A(x) \, = \, x^2 \left[ \log\left(\frac{x^2}{\mu^2}\right)\,-\,1\right]
\ee
and
\be
B(x,y,p^2) \, = \, \int^1_0 \,dt\, \log \left[ \frac{t\,x^2 \,+\, (1 - t)\,y^2 \,-\,
t(1 - t) p^2}{\mu^2}\right].
\label{eq:B}
\ee
Further, the gauge contributions are given by
\ba
m^2_{h_{1,G}} &=& \frac{g^2}{2 c^2_W}\,B_{SV}(m_{G_0},m_{Z_0},p^2) \,+\,
g^2\,B_{SV}(m_{G^\pm_0},m_{W_0},p^2)
\vspace{0.4cm} \nonumber \\
 & + & \frac{g^2}{c^2_W}\,m^2_{Z_0}\,B_{VV}(m_{Z_0},m_{Z_0},p^2) \,+\,
 2\,g^2\,m^2_{W_0}\,B_{VV}(m_{W_0},m_{W_0},p^2)
\vspace{0.4cm} \nonumber \\
 & + & \frac{3 g^2}{2 c^2_W}\,A(m_{Z_0}) \, + \, 3 \,g^2\,A(m_{W_0})\,,
\ea
where the functions $B_{SV}$ and $B_{VV}$ correspond to self-energy diagrams with,
respectively, one and two internal gauge boson lines. They are given by (in $B_{SV}$ $x$ stands for the
scalar's mass squared, and $y$ for the squared mass of the gauge boson)
\ba
B_{SV}(x,y,p^2) & = & \frac{(x^2 - y^2 - p^2)^2\,-\,4\,y^2\,p^2}{y^2}\,B(x,y,p^2) \,-\,
\frac{(x^2 - p^2)^2}{y^2}\,B(x,0,p^2)\,+\, A(x) \nonumber \\
 & + &
\frac{x^2 - y^2 - p^2}{y^2}\, A(y),
\label{eq:BSV}
\vspace{0.4cm}  \\
B_{VV}(x,y,p^2) & = & \frac{A(x)}{4\,x^2} \,+\,\frac{A(y)}{4\,y^2} \,+\,
2\,B(x,y,p^2) \,+\,
\frac{(x^2 + y^2 - p^2)^2}{4x^2 y^2} \, B(x,y,p^2)
\vspace{0.4cm} \nonumber \\
 & - &  \frac{(x^2 - p^2)^2}{4x^2 y^2}\,B(x,0,p^2) \,-\, \frac{(y^2 - p^2)^2}{4x^2 y^2}\,B(0,y,p^2)
 \,+\, \frac{p^4}{4x^2 y^2} \, B(0,0,p^2).
 \label{eq:BVV}
\ea
These functions are written assuming non-zero gauge boson masses. For massless gauge bosons
infrared divergences may in principle arise, and the limits of zero masses need to be handled with special care.
Those special cases
\cite{Martin:2003qz,Martin:2003it})
may be found in Appendix~\ref{ap:specialB}.

Finally, the fermionic contributions are given by
\ba
m^2_{h_{1,F}} &=& -\,6\,\lambda_t^2\,\left[\frac{}{} (4\,m_{t_0}^2 - p^2)\,B(m_{t_0},m_{t_0},p^2) \, +\,
2 \, A(m_{t_0}) \right]
\vspace{0.4cm} \nonumber \\
& & -\,6\,\lambda_b^2\,\left[\frac{}{} (4\,m_{b_0}^2 - p^2)\,B(m_{b_0},m_{b_0},p^2) \, +\,
2 \, A(m_{b_0}) \right]
\vspace{0.4cm} \nonumber \\
& & -\,2\,\lambda_\tau^2\,\left[\frac{}{}(4\,m_{\tau_0}^2 - p^2)\,B(m_{\tau_0},m_{\tau_0},p^2) \, +\,
2 \, A(m_{\tau_0}) \right].
\label{eq:mhf}
\ea
The one-loop contribution~\eqref{eq:h1} to the $h$ mass must be computed at the physical value of the mass,
meaning we must take $p^2 \,=\, m^2_h(physical) \,=\,(125$ GeV)$^2$.

\subsection*{ For the inert scalar $H$:}
\be
m^2_H \,=\,m^2_{H_0} \,+\,\frac{1}{32\pi^2}\,m^2_{H_1},
\label{eq:mH}
\ee
with $m^2_{H_0}$ given by eq.~\eqref{eq:mH0} and we have
\be
m^2_{H_1} \;=\; \mbox{Re}(m^2_{H_{1,S}} \,+\, m^2_{H_{1,G}} )\,.
\ee
There are no fermionic contributions to the $H$ mass in the IDM, since the doublet
$\Phi_2$ does not couple to fermions. We then obtain
\ba
m^2_{H_{1,S}} &=& \bar{\lambda}_{345} A(m_{G_0}) \,+\,2\lambda_3 A(m_{G^\pm_0}) \,+\,\lambda_{345} A(m_{h_0})
\,+\,
3\lambda_2 A(m_{H_0}) \,+\,\lambda_2 A(m_{A_0}) \,+\, 2\lambda_2 A(m_{H^\pm_0})
\vspace{0.4cm} \nonumber \\
  & + & 2\lambda_{345}^2 \,v_1^2 \,B(m_{H_0},m_{h_0},p^2) \,+\, 2\lambda_5^2 \,v_1^2 \,B(m_{A_0},m_{G_0},p^2)\nonumber\\[.1cm]
  &+& (\lambda_4 + \lambda_5)^2 \,v_1^2 \,B(m_{H^\pm_0},m_{G^\pm_0},p^2) \,,
  \label{eq:H1}
\ea
and
\ba
m^2_{H_{1,G}} &=& \frac{g^2}{2 c^2_W}\,B_{SV}(m_{A_0},m_{Z_0},p^2) \,+\,
g^2\,B_{SV}(m_{H^\pm_0},m_{W_0},p^2)\nonumber\\
&+& \frac{3 g^2}{2 c^2_W}\,A(m_{Z_0}) \, + \, 3 \,g^2\,A(m_{W_0})\,.
\label{eq:mHG}
\ea
Once more, all functions must be calculated at  $p^2 \,=\, m^2_H(physical)$. This can be ensured in
one of two ways: either we choose values for the physical $H$ mass, on a given scan of the inert
model parameter space (and find the values of the couplings $\lambda_i$, $m_{jj}^2$ that produce
that physical mass); or (if we have already specified all parameters of the potential) we find,
via an iterative procedure, the value of $p^2$ which satisfies the equation $p^2 = m^2_H(p^2)$,
with $m^2_H(p^2)$ taken from eq.~\eqref{eq:mH}.

\subsection*{For the inert scalar $A$:}

\be
m^2_A \,=\,m^2_{A_0} \,+\,\frac{1}{32\pi^2}\,m^2_{A_1},
\label{eq:mA}
\ee
with $m^2_{A_0}$ given by eq.~\eqref{eq:mA0} and
\be
m^2_{A_1} \;=\; \mbox{Re}(m^2_{A_{1,S}} \,+\, m^2_{A_{1,G}} )\,.
\ee
Once more there are no fermionic contributions to this mass, for the same reason as presented for
the $H$ mass. We then have
\ba
m^2_{A_{1,S}}  &=& \lambda_{345} A(m_{G_0}) \,+\,2\lambda_3 A(m_{G^\pm_0}) \,+\,\bar{\lambda}_{345} A(m_{h_0})
\,+\,
\lambda_2 A(m_{H_0}) \,+\,3\lambda_2 A(m_{A_0}) \,+\, 2\lambda_2 A(m_{H^\pm_0})
\vspace{0.2cm} \nonumber \\
  & + & 2\bar{\lambda}_{345}^2 \, v_1^2 \,B(m_{A_0},m_{h_0},p^2) \,+\,
  2\lambda_5^2 \, v_1^2 \,B(m_{H_0},m_{G_0},p^2)\nonumber\\[.1cm]
  &+& (\lambda_4 - \lambda_5)^2 \,v_1^2 \,B(m_{H^\pm_0},m_{G^\pm_0},p^2) \,,
  \label{eq:A1}
\ea
and
\ba
m^2_{A_{1,G}} &=& \frac{g^2}{2 c^2_W}\,B_{SV}(m_{H_0},m_{Z_0},p^2) \,+\,
g^2\,B_{SV}(m_{H^\pm_0},m_{W_0},p^2) \nonumber\\
&+&
 \frac{3 g^2}{2 c^2_W}\,A(m_{Z_0}) \, + \, 3 \,g^2\,A(m_{W_0})\,.
 \label{eq:mAG}
\ea
Again, all functions are computed with $p^2 \,=\, m^2_A(physical)$.

\subsection*{For the charged scalar $H^{\pm}$:}

\be
m^2_{H^\pm} \,=\,m^2_{H^\pm_0} \,+\,\frac{1}{32\pi^2}\,m^2_{H^\pm_1},
\label{eq:mch}
\ee
with $m^2_{H^\pm_0}$ given by eq.~\eqref{eq:mHch0} and
\be
m^2_{H^\pm_1} \;=\; \mbox{Real}(m^2_{H^\pm_{1,S}} \,+\, m^2_{H^\pm_{1,G}})\,,
\ee
and given $H^\pm$ does not couple to fermions, there are no fermionic contributions.
We then have
\ba
m^2_{H^\pm_{1,S}} &=& \lambda_3 A(m_{G_0}) \,+\,2 (\lambda_3 + \lambda_4) A(m_{G^\pm_0}) \,+\,
\lambda_3 A(m_{h_0}) \,+\,
\lambda_2 A(m_{H_0}) \,+\,\lambda_2 A(m_{A_0}) \,+\, 4\lambda_2 A(m_{H^\pm_0})
\vspace{0.2cm} \nonumber \\
  & + & 2\lambda_3^2 \, v_1^2 \,B(m_{h_0},m_{H^\pm_0},p^2)
  \,+\, \frac{1}{2}(\lambda_4 - \lambda_5)^2 \, v_1^2 \,B(m_{A_0},m_{G^\pm_0},p^2) \nonumber\\
  &+&
   \frac{1}{2}(\lambda_4 + \lambda_5)^2 \,v_1^2 \,B(m_{H_0},m_{G^\pm_0},p^2) \,,
  \label{eq:Hch1}
\ea
\ba
m^2_{H^\pm_{1,G}} &=& 2 e^2B_{SV}(m_{H^\pm_0},0,p^2) \,+\,
2 e^2\,\cot^2_{2W}B_{SV}(m_{H^\pm_0},m_{Z_0},p^2)\nonumber\\
&+&
\frac{g^2}{2}\left[B_{SV}(m_{H_0},m_{W_0},p^2)\,+\,
B_{SV}(m_{A_0},m_{W_0},p^2)\right]
\vspace{0.4cm} \nonumber \\[.1cm]
 & + & 6 e^2\,\cot^2_{2W}\,\,A(m_{Z_0}) \, + \, 3\, g^2 \,A(m_{W_0})\,.
 \label{eq:mchG}
\ea
All functions must be computed with $p^2 \,=\, m^2_{H^\pm}(physical)$.

\subsection*{For the neutral Goldstone $G$:}

\be
m^2_G \,=\,m^2_{G_0} \,+\,\frac{1}{32\pi^2}\,m^2_{G_1},
\label{eq:mG1}
\ee
with $m^2_{G_0}$ given by eq.~\eqref{eq:mG0} and
\be
m^2_{G_1} \;=\; \mbox{Re}(m^2_{G_{1,S}} \,+\, m^2_{G_{1,G}} \,+\, m^2_{G_{1,F}})\, .
\ee
Since the neutral Goldstone corresponds to the imaginary part of the neutral component of
the $\Phi_1$ doublet, it does couple to fermions, and therefore there are fermionic
contributions in the expression above. We then have, for the scalar contributions,
\ba
m^2_{G_{1,S}} &=& 3\lambda_1 A(m_{G_0}) \,+\,2\lambda_1 A(m_{G^\pm_0}) \,+\,\lambda_1 A(m_{h_0})
\,+\,\bar{\lambda}_{345} A(m_{H_0}) \,+\,\lambda_{345} A(m_{A_0}) \,+\, 2\lambda_3 A(m_{H^\pm_0})
\vspace{0.2cm} \nonumber \\
  & + & 2 \lambda_1^2 \, v_1^2 \,B(m_{G_0},m_{h_0},p^2) \,+\,
  2\lambda_5^2 \, v_1^2 \,B(m_{H_0},m_{A_0},p^2) \,,
\ea
for the gauge contributions,
\ba
m^2_{G_{1,G}} &=& \frac{g^2}{2 c^2_W}\,B_{SV}(m_{h_0},m_{Z_0},p^2)\,+\,g^2\,B_{SV}(m_{G^\pm_0},m_{W_0},p^2)\nonumber\\
&+&
\frac{3 g^2}{2 c^2_W}\,A(m_{Z_0}) \,+\, 3\, g^2\,A(m_{W_0}) \, ,
\ea
and for the fermionic ones,
\ba
m^2_{G_{1,F}} &=& 6 \lambda_t^2\,\left[p^2\,B(m_{t_0},m_{t_0},p^2) \,-2\,\,A(m_{t_0})\right] \, +\,
6\lambda_b^2\,\left[p^2\,B(m_{b_0},m_{b_0},p^2) \,-2\,\,A(m_{b_0})\right]
\vspace{0.2cm} \nonumber \\
  & + & 2\lambda_{\tau}^2\,\left[p^2\,B(m_{\tau_0},m_{\tau_0},p^2) \,-2\,\,A(m_{\tau_0})\right].
\label{eq:G1f}
\ea

Since the Goldstone bosons have zero physical mass, the expressions must be
evaluated at $p^2 = 0$. This simplifies considerably the results (for instance, all $B_{SV}$
functions are identically zero at zero external momentum, and there are further simplifications
with the $B$ functions). A direct calculation at the one-loop minimum then confirms that the
Goldstone bosons are massless --- the one-loop minimum condition ensures that the expression
in~\eqref{eq:mG1} is equal to zero.

\subsection*{For the charged Goldstone $G^\pm$:}

\be
m^2_{G^\pm} \,=\,m^2_{G^\pm_0} \,+\,\frac{1}{32\pi^2}\,m^2_{G^\pm_1},
\label{eq:mGch}
\ee
with $m^2_{G^\pm_0}$ given by eq.~\eqref{eq:mGch0} and
\be
m^2_{G^\pm_1} \;=\; \mbox{Re}(m^2_{G^\pm_{1,S}} \,+\, m^2_{G^\pm_{1,G}} \,+\, m^2_{G^\pm_{1,F}}),
\ee
with
\ba
m^2_{G^\pm_{1,S}} &=& \lambda_1 A(m_{G_0}) \,+\,4\lambda_1 A(m_{G^\pm_0}) \,+\,\lambda_1 A(m_{h_0})
\,+\,\lambda_3 A(m_{H_0}) \,+\,\lambda_3 A(m_{A_0}) \,+\, 2(\lambda_3 + \lambda_4) A(m_{H^\pm_0})
\vspace{0.2cm} \nonumber \\
  & + & 2 \lambda_1^2 \, v_1^2 \,B(m_{G_0^{\pm}},m_{h_0},p^2) \,+\,
  \frac{1}{2} (\lambda_4 - \lambda_5)^2 \, v_1^2\, B(m_{A_0},m_{H^\pm_0},p^2) \nonumber\\[.1cm]
  &+&
  \frac{1}{2} (\lambda_4 + \lambda_5)^2 \, v_1^2\, B(m_{H_0},m_{H^\pm_0},p^2) \,,
\ea
\ba
m^2_{G^\pm_{1,G}} &=& \frac{g^2 c^2_{2W}}{2c^2_W}\,B_{SV}(m_{G^\pm_0},m_{Z_0},p^2)\,+\,
\frac{g^2}{2}\,\left[B_{SV}(m_{G_0},m_{W_0},p^2) + B_{SV}(m_{h_0},m_{W_0},p^2)\right]\nonumber\\
&+&
2\,e^2 \,B_{SV}(m_{G^\pm_0},0,p^2)+  2 e^2 m^2_{W_0}\,\left[B_{VV}(0,m_{W_0},p^2)\,+\,\frac{s_W^2}{c_W^2}\,B_{VV}(m_{Z_0},m_{W_0},p^2) \right]\nonumber\\
&+ & \frac{6 e^2 c^2_{2W}}{s^2_{2W}}\,A(m_{Z_0}) \,+\, 3\, g^2\,A(m_{W_0}) \, ,
\ea%

and
\ba
m^2_{G^\pm_{1,F}} &=& 6\,\left\{ \left[ (\lambda_t^2 + \lambda_b^2)\,p^2 \,-\,
\frac{1}{2}(\lambda_t^2 - \lambda_b^2)^2 v_1^2\right]\,B(m_{t_0},m_{b_0},p^2)\, - \,
(\lambda_t^2 + \lambda_b^2)\,\left[A(m_{t_0})\,+\,A(m_{b_0})\right]\right\}
\vspace{0.2cm} \nonumber \\
 & + & 2\,\lambda_\tau^2\left[\left(p^2 \,-\, \frac{1}{2}\lambda_\tau^2 v_1^2
\right) \,B(m_{\tau_0},m_{\tau_0},p^2)\, - \,A(m_{\tau_0})\right]\,.
\label{eq:mGchf}
\ea
Once again, all $B$ functions must be computed with
$p^2 \,=\, 0$. The expressions obtained in that regime are identical to those obtained for
the neutral Goldstone in the same regime,
and again yield massless charged Goldstone bosons at the one-loop minimum.

\subsection{Inert-like vacua}

Inert-like vacua have $\langle r_2\rangle = v_2$ ($v_2$, of course, is in general not equal to 246 GeV), and other vevs equal to zero. Since the $\Phi_2$ doublet does not couple
to fermions, in this vacuum the fermions are massless. Notice, however, that the lagrangian of the model
includes the Yukawa couplings of $\Phi_1$ (eq.~\eqref{eq:lyuk}), so there will be Yukawa contributions
to the one-loop masses at the inert-like vacuum, for the scalars pertaining to the doublet $\Phi_1$.\footnote{This may be thought of in the following manner: in the inert-like minimum, all fermions are massless,
and there
is an interaction lagrangian between $\Phi_1$ and those fermions. The scalars arising from $\Phi_1$ therefore
couple, and decay, to massless fermions, and are thus unstable. This is an added reason to consider the
inert-like vacuum unphysical, and to avoid it at all costs.} The expressions presented so far for
inert vacua can be easily adapted for inert-like vacua: first, in all vevs and scalar couplings' indices,
the swap $1\leftrightarrow 2$ must be performed; then, case by case, one has:
\begin{itemize}
\item The tree-level masses in the inert-like vacuum can be read from eqs.~\eqref{eq:mG0}--\eqref{eq:mfG}
by setting all fermion masses to zero.
\item The minimisation condition which determines the existence of an inert-like stationary point can be
read from eq.~\eqref{eq:dVt}, by setting all fermion contributions to zero.
\item The one-loop mass of the $r_2$ field (which will be the ``h'' scalar of the inert-like vacuum) can be
read from eqs.~\eqref{eq:mh}--\eqref{eq:mhf}, by setting all fermion contributions to zero.
\item The one-loop mass of the $i_2$ field (which will be the neutral Goldstone of the inert-like vacuum)
can be read from eqs.~\eqref{eq:mG1}--\eqref{eq:G1f}, by setting all fermion contributions to zero.
\item The one-loop mass of the $c_3$, $c_4$ fields (which will be the charged Goldstones of the inert-like
vacuum) can be read from eqs.~\eqref{eq:mGch}--\eqref{eq:mGchf}, by setting all fermion contributions to
zero.\footnote{Technically the charged Goldstones correspond to $(c_3 \pm i c_4)/\sqrt{2}$, not
$c_3$ and $c_4$ themselves. Likewise, the
charged scalars correspond to $(c_1 \pm i c_2)/\sqrt{2}$, not $c_1$ and $c_2$.}
\item The one-loop mass of the $r_1$ field (which will be the ``H'' scalar of the inert-like vacuum) can be
read from eqs.~\eqref{eq:mH}--\eqref{eq:mHG}. But since $r_1$ has Yukawa couplings to fermions, there is a
fermionic contribution to this mass. It is given by eq.~\eqref{eq:mhf} setting all fermion masses (but not
the Yukawa couplings) to zero, and is thus equal to
\be
\frac{1}{32\pi^2}\,(6\lambda_t^2 + 6 \lambda_b^2 + 2 \lambda_\tau^2)\,p^2\,B(0,0,p^2)\,.
\label{eq:c0f}
\ee
\item
The one-loop mass of the $i_1$ field (which will be the ``A'' scalar 
 of the inert-like vacuum) can be
read from eqs.~\eqref{eq:mA}--\eqref{eq:mAG}. Additionally, since $i_1$ has Yukawa couplings to fermions, there is a fermionic
contribution to this mass, given by eq.~\eqref{eq:G1f} setting all fermion masses to zero. It is equal
to eq.~\eqref{eq:c0f}.
\item
The one-loop mass for the $c_1$, $c_2$ fields (which will be the ``$H^\pm$'' charged scalar of
the inert-like vacuum) can be read from eqs.~\eqref{eq:mch}--\eqref{eq:mchG}. Again, there is a
fermionic contribution to this mass, given by eq.~\eqref{eq:mGchf} setting all fermion masses
(but not the Yukawa couplings) to zero. It is equal to eq.~\eqref{eq:c0f}.
\end{itemize}

\section{Comparison of depths of inert and inert-like minima}
\label{sec:comp}

With the expressions presented in the previous section, we are now capable of performing a parameter space
scan of the 2HDM parameter space, looking for coexisting inert and inert-like minima. Briefly,
our procedure was as follows:
\begin{itemize}
\item The minimisation conditions of the potential, eq.~\eqref{eq:dVt} and their analogue for the
inert-like minimum, were required to have {\em simultaneous} solutions. We force $v_1 = v =$ 246 GeV, so that the inert
vacuum can correspond to the real world, requiring at the same time that the $h$ scalar in that vacuum
has a mass of about 125 GeV. We find a numerical solution of the minimisation conditions obtaining a set of
the quadratic couplings $m^2_{ii}$, the quartic couplings $\lambda_j$ and the value of the vev
at the inert-like vacuum, $v_2$.
\item Having discovered a set of parameters for which the two stationary points exist simultaneously,
we proceed to verify whether they are minima --- to do so we compute the one-loop self energies of
all scalars at both vacua
and verify whether they are positive. If they are, we have coexisting
minima for this choice of parameters, which is stored. If this test fails, the minimisation procedure
is once again performed.
\item We also verify that in each of the minima the Goldstones' masses are zero. This involves a
subtlety: since we are performing a numerical computation, the expressions used for the Goldstone masses
do not give exactly zero, due to precision limitations. Nonetheless, they produce extremely low masses (less than
$10^{-6}$, versus the mass of the other scalars, hundreds of GeV). In practice this means that
we are computing these masses not {\em exactly} at the minima, but rather close to them --- and therefore one
must require that the Goldstones' masses are {\em small} but {\em always positive}. If they were allowed to be
negative (and small in absolute value) the stationary point would in fact be a saddle point.
\item As was explained above, the procedure to find the physical masses of the scalars is twofold: for the
inert minimum, the $h$ scalar is forced to have the observed mass of 125 GeV;\footnote{In fact we allowed for
an interval of 1 GeV around this central value} for the remaining scalars in both minima, the physical
masses are computed via the iterative procedure described just after eq.~\eqref{eq:mHG}.
\item Once the existence of both minima has been confirmed, a final check is performed, to make sure
that the values of the quartic couplings found conform to the bounded from below conditions of
eqs.~\eqref{eq:bfb} and the unitarity bounds of~\cite{Kanemura:1993hm,Akeroyd:2000wc, Swiezewska:2012yuk},
and are sufficiently small (absolute values below $4\pi$) so
as to ensure that the model is perturbative.
\item At this stage the value of the one-loop potential is calculated at each of the minima, and the
scan continues searching for a different set of parameters.
\item We also take a conservative approach and only accept one-loop minima for which all tree-level
squared masses are positive. This might be thought of as looking for one-loop minima around tree-level
ones, and ensures that no complex components to the one-loop potential appear.
\end{itemize}

Before we proceed with the results of our analysis, a word on the choice of the renormalisation
scale $\mu$. This choice is of course arbitrary --- up to two-loop effects, neither the effective potential,
nor the vevs nor the physical masses depend on $\mu$.
In fact this is not 100\% accurate: such as
it is written in eq.~\eqref{eq:Vt}, the one-loop effective potential {\em does} depend on the renormalisation
scale, as it is lacking a field-independent term, $\Omega$, which compensates said dependence~\cite{Ford:1992mv}.
However, being field-independent, $\Omega$ has the exact same value at both minima, so the difference
in the values of the potential at both minima, $V_I - V_{IL}$, is a renormalization scale-independent
quantity. This is the same argument that was used, for the MSSM potential, in
refs.~\cite{Ferreira:2000hg,Ferreira:2001tk}, using the results of~\cite{Ford:1992mv}. However, the results
of~\cite{Ford:1992mv} are valid for a generic multi-scalar theory, so they also apply to our study of the IDM.

The only guiding principle behind the choice of value of $\mu$ is to attempt to render all (or most) of
the logarithms present in the calculations ``small'' --- thus a choice of $\mu$ of the order of the masses
present in the theory is advocated~\cite{Gamberini:1989jw}. However, unlike what is usually
considered~\cite{Casas:1995pd}, there are strong arguments~\cite{Ferreira:2000hg,Ferreira:2001tk} for
comparing the value of the potential at both minima {\em at the same renormalisation scale}.

Our goal is to study the differences between tree-level and one-loop comparison of the values of
the potential at the two minima. Since the formalism adopted guarantees that our conclusions are
independent of the value of $\mu$, any choice is acceptable, and we set $\mu = 200$ GeV, which we
found is close enough to the typical masses discovered at these minima. If one wished to verify our
conclusions at another scale $\mu^\prime$, all couplings in the model would have to be ran from $\mu$
to $\mu^\prime$ by means of the IDM $\beta$-functions (see, for
instance,~\cite{Haber:1993an,Ferreira:2009,Ferreira:2015rha}). We kept the renormalisation scale always
fixed, however, and therefore all couplings are evaluated/obtained at that scale. The top quark Yukawa
coupling, for instance, is obtained from the one-loop relationship between the pole mass and
the running one, {\em i.e.}
\be
m_t^{(pole)}\,=\,m_t(\mu)\,\left[1 \,+\,\left(5\,-\,3\,\log\left(\frac{m_t^2(\mu)}{\mu^2}\right)\,
\right)\frac{\alpha_S(\mu)}{3\,\pi}\right]\,,
\label{eq:mtp}
\ee
where $m_t(\mu)\,=\,\lambda_t(\mu)\,v_1/\sqrt{2}$ is the top quark running mass, computed from
eq.~\eqref{eq:mf0} with the
running Yukawa coupling $\lambda_t(\mu)$. In eq.~\eqref{eq:mtp} we only kept the one-loop strong interaction
contributions, proportional to the strong gauge coupling $\alpha_S$, since the electroweak corrections
to $m_t^{(pole)}$ are much smaller.\footnote{For consistency, eq.~\eqref{eq:mtp} was obtained with the DRED
regularisation scheme. In the more frequent dimensional regularisation scheme, the factor of ``5'' would become
a factor of ``4''.}

\subsection{Scalar contributions only}

To begin with, we are going to consider a toy-model: a 2HDM {\em only with scalars}. In other words, a theory
containing no fermions, and for which the gauge symmetry $SU(3)\times SU(2) \times U(1)$ is {\em global},
and therefore there are no gauge bosons. Our intention is simple: as we will shortly see, the tree-level
predictions concerning the relative depths of the inert and inert-like minima differ from those obtained
using the one-loop potential and masses. We wish to ascertain exactly {\em where} those differences
come from --- from the scalar, gauge or fermionic sectors.

\begin{figure}[ht]
\centering
\includegraphics[height=8cm,angle=0]{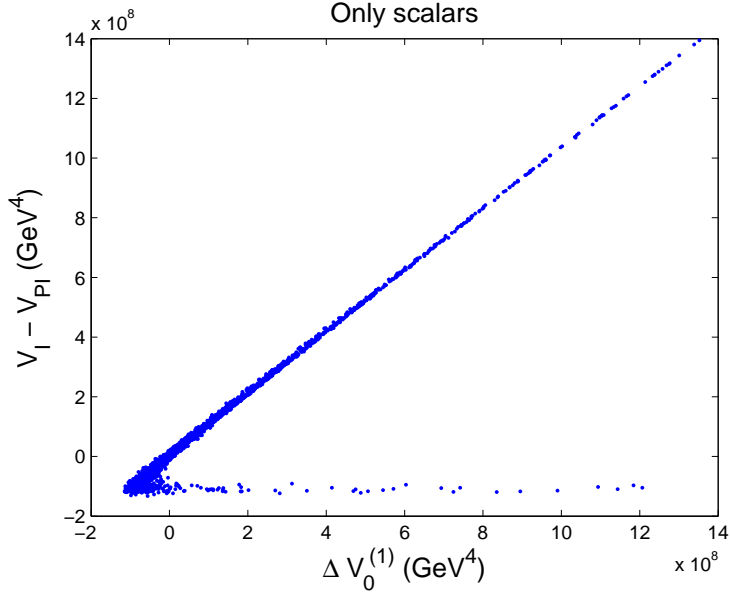}
\caption{One-loop computed difference in inert and inert-like minima depths ($V_I \,-\,V_{PI}$) {\em versus} the
tree-level expected depth difference given by eq.~\eqref{eq:difV01}.}
\label{fig:dV1_S}
\end{figure}

We are therefore in a theory without gauge bosons or fermions, and therefore one must simply set to
zero the gauge and fermionic contributions to the effective potential, eqs.~\eqref{eq:Vt}--\eqref{eq:V1}, to
its derivatives, eq.~\eqref{eq:dVt}, and to the scalar masses, eqs.~\eqref{eq:mh}--\eqref{eq:mGchf}. In
eqs.~\eqref{eq:difV01}--\eqref{eq:difV02} we presented the analytical expressions obtained
at tree level relating the difference in the depths of the potential at the inert and inert-like
minima. We are now in condition to investigate whether they hold at the one-loop level. Let us call
$\Delta V_0^{(1)}$ the tree-level expression of eq.~\eqref{eq:difV01}, in which the
difference in the values of the potential is obtained directly in terms of the parameters of the potential.
In fig.~\ref{fig:dV1_S} we plot the difference in the value of the potential at one-loop in both minima
{\em versus} the tree-level prediction $\Delta V_0^{(1)}$ (computed with the parameters which satisfy the one-loop
minimisation). As can easily be seen, there are deviations
between the tree-level formula and the one-loop results. Nonetheless, even though the tree-level formula
does not give exactly the same values as the one-loop results, it is clear that, for most of the parameter
space, eq.~\eqref{eq:difV01} is a very good approximation to the one-loop potential depth difference.

However, there is a very small subset of points (of the order of 2.8\% of all sets of points in the
parameter space simulated\footnote{This percentage does not have any particular physical meaning,
given that we are not scanning the {\em whole} of the parameter space, neither it is ensured that each
area of it is being scanned with equal weights. Nonetheless, the numbers are informative, in the sense that
they are {\em small}, obtained in a ``blind search'' for minima, in which no specific region of parameter
space was singled out as special.})
for which the tree-level formula of eq.~\eqref{eq:difV01} predicts that the global minimum of the theory
(inert or inert-like) is {\em the reverse} of what one obtains with the one-loop effective potential.
This only occurs in two cases: the first corresponds to some choices of parameters for which $V_I \,-\,V_{IL}$
is ``close to zero'' and both minima are close to degenerate. It is reassuring that this {\em inversion} of
minima occurs when they are nearly degenerate --- it is a sign that the difference between tree-level
and one-loop results is not being caused by a breakdown of perturbation theory, but rather is an
understandable loop effect over a quantity which, at tree level, is small.

\begin{figure}[ht]
\centering
\includegraphics[height=8cm,angle=0]{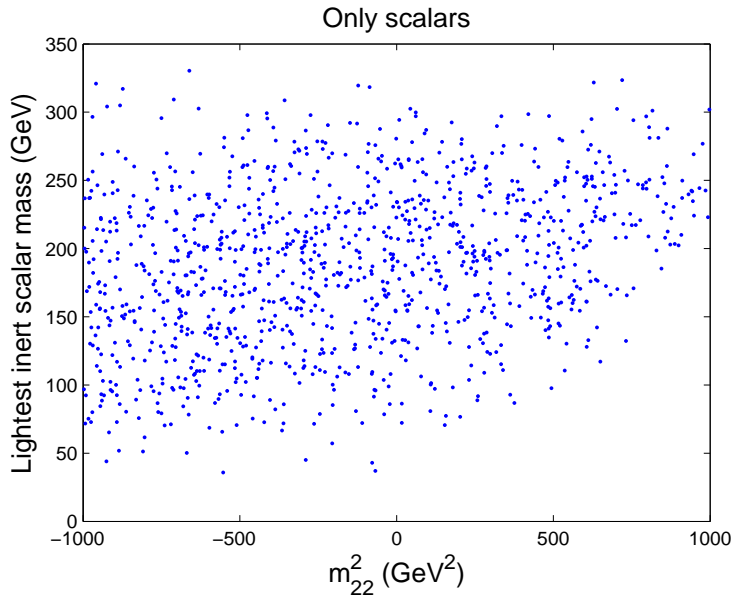}
\caption{The parameter $m^2_{22}$ {\em versus} the lightest neutral inert scalar at the
inert minimum. For all points shown, one-loop inert and inert-like minima coexist. At tree level, all
points with $m^2_{22} > 0$ would not be present.}
\label{fig:m22_S}
\end{figure}
The second situation for which minima inversion may occur corresponds to the strange points which deviate
from the central band, in the lowest part of fig.~\ref{fig:dV1_S}. These correspond to a region of
parameter space for which  $m^2_{22} > 0$: such a condition was not allowed, for coexisting minima,
at tree level. In other words, the one-loop minimisation allows the coexistence of inert and inert-like
minima for a vaster region of parameter space than what would be expected from a tree-level analysis.
This may be appreciated from fig.~\ref{fig:m22_S}, where we plot the value of the $m^2_{22}$ parameter
{\em versus} the mass of the lightest neutral inert scalar (at the inert minimum).
It is plain to see that, even in this situation for which  inert and inert-like minima coexist in the model,
plenty of points with $m^2_{22} > 0$ are found. We again stress that this was an impossibility at
tree level. And it does {\em not} correspond to a breakdown of perturbation theory --- rather, the one-loop
procedure opens up new regions of parameter space, which seemed impossible to reach using a tree-level
analysis alone.

\begin{figure}[ht]
\centering
\includegraphics[height=8cm,angle=0]{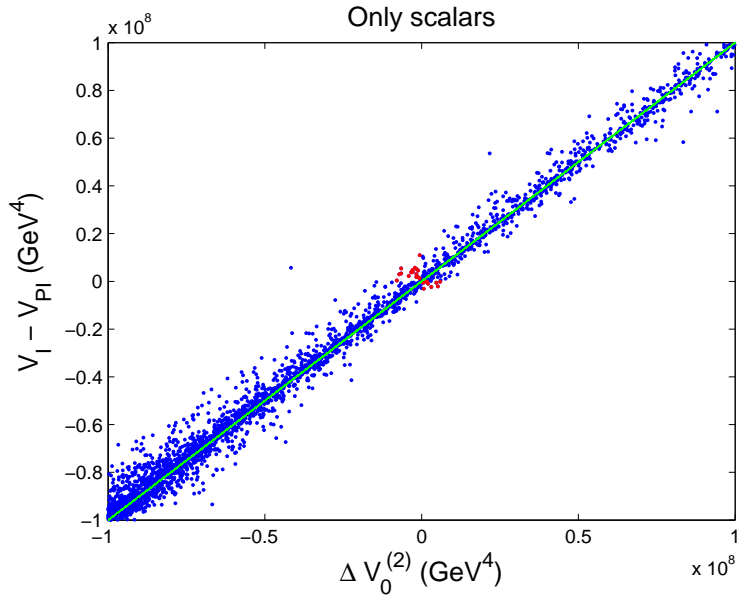}
\caption{One-loop computed difference in inert and inert-like minima depths ($V_I \,-\,V_{PI}$)
{\em versus} the tree-level expected depth difference given by eq.~\eqref{eq:difV02}. Notice that,
compared to fig.~\ref{fig:dV1_S}, we are only showing the region closer to near-degenerate minima.
The green line corresponds to the tree-level prediction. The red points are those for which the
one-loop minimum has the opposite type of what is predicted
by the tree-level analysis.}
\label{fig:dV2_S}
\end{figure}

There is another point which should be stressed --- the equality between the formulae of
eq.~\eqref{eq:difV01} and eq.~\eqref{eq:difV02} is only valid at tree level. At one-loop,
it no longer holds. In fact, let $\Delta V_0^{(2)}$ be the formula of eq.~\eqref{eq:difV02} --- it
depends on the vevs and charged scalar masses at each minimum. Let us then compare the difference
in depths of the one-loop potential at both minima with $\Delta V_0^{(2)}$, {\em computed using
the one-loop vevs and one-loop charged masses}. What we conclude is that eq.~\eqref{eq:difV02}
is {\em also} not an exact description for the relative depth of the minima. The number of
inversions of one-loop minima {\em vis a vis} what is predicted by $\Delta V_0^{(2)}$ is slightly
smaller than earlier --- of the order of 0.6\% of the simulated points, shown in red in fig.~\ref{fig:dV2_S},
but a curious detail
emerges: these points for which one finds that the one-loop global minimum is inert/inert-like
compared to the tree-level prediction of an inert-like/inert minimum are not entirely a subset
of the 2.8\% points for which this occurred when using $\Delta V_0^{(1)}$. Rather, there are
some new points. Therefore, the formulae obtained for the relative potential depth not only
do not give exact results at the one-loop level, they also do not coincide in their predictions.
However, as may be appreciated from fig.~\ref{fig:dV2_S}, $\Delta V_0^{(2)}$ is also an
excellent approximation to the one-loop potential difference --- in fact, a better one, since the $m^2_{22} > 0$
points, using the formula for $\Delta V_0^{(2)}$, do not deviate overmuch from the central band.
The green line shown in the
plot shows what one would obtain if $\Delta V_0^{(2)}$ gave results {\em exactly equal} to
the one-loop potential difference. The dispersion of points that is verified around that central
line shows that the one-loop results do differ, though not by much, from the tree-level derived
formula. The red points are those for which an inversion of minima occurs.

All told, these results allow us to conclude that:

\begin{quote}
\em The tree-level results for the relative depths of the inert and inert-like minima
may  wrongly predict what the global minimum of the model is. The one-loop allowed
parameter space of the model is larger than that which the tree-level analysis specifies.
\end{quote}

\subsection{Scalar and gauge contributions only}

We have, in the previous subsection, already established that the tree-level
predictions for the relative depths of the inert and inert-like minima are not entirely respected
by the one-loop potential. As one goes to a one-loop analysis, in which all sectors of the theory
(not only the scalar one) contribute to the potential, one might expect that eventual differences in the
vacuum structure of the model might be due to the new sectors --- gauge or fermion. But what the analysis
we presented so far has shown is that the differences found between tree-level and one-loop results
can emerge {\em as a consequence
of the scalar sector alone}. It is however interesting to verify whether the inclusion of the gauge sector
makes any changes in the results shown in the previous section. To that end, we consider a toy model
with {\em local} gauge symmetry $SU(3)\times SU(2) \times U(1)$, but without any fermions. The contributions to
the potential, eqs.~\eqref{eq:Vt}--\eqref{eq:V1}, to its derivatives, eq.~\eqref{eq:dVt}, and to the scalar
masses, eqs.~\eqref{eq:mh}--\eqref{eq:mGchf}, are therefore obtained simply by setting the fermionic
contributions to zero in all of those expressions. We then proceed with a new scan of parameter space
following the procedure described above and compare the potentials at the inert and inert-like minima.

The conclusion is simple: there are no substantial differences from what we had already concluded
from the analysis of the scalar sector alone. Once more, the one-loop analysis reveals a vaster
parameter space than what is allowed with the tree-level potential; the formulae of
eqs.~\eqref{eq:difV01} and~\eqref{eq:difV02} are a good approximation to the one-loop
results, the latter doing a better job than the former; and again there is a small number
of points for which the one-loop global minimum is of the opposite nature than what one expected
from the tree-level results. The equivalent figures to figs.~\ref{fig:dV1_S} to~\ref{fig:dV2_S}
are quite similar to them, therefore we will not display them in this subsection.

Why this similitude? In fact, it was to be expected --- at both the inert and inert-like minima, the
gauge sector has contributions which, though they may differ {\em numerically} (because the vevs
are different in both minima, thus the masses of the W and Z bosons are also different), are
{\em qualitatively} identical. In both minima, the gauge sector contributes to the masses of all
scalars, to the potential and its derivatives, in equivalent manners. It is therefore natural
that the gauge sector by itself does not introduce anything qualitatively new in the comparison
of the inert and inert-like minima. But as we are about to see, that will no longer be the
case when the fermions are finally taken into account.

\subsection{Scalar, gauge and fermionic contributions}

Having established that differences in the relative depths of the potential at inert and inert-like
minima arise in the one-loop minimisation due to the scalar sector alone; and that the gauge sector
does not affect those conclusions, since it contributes in the same manner to the physics of both
vacua; we now turn to the effect of the fermions on the one-loop minima. And here we should expect
that the fermions will have a distinct contribution to the inert minimum and to the inert-like one. In fact,
at the inert minimum the fermions contribute: (a) to the value of the potential, (b) to its first
derivatives and (c) to the masses of the Higgs boson 
and the Goldstone bosons. Conversely, other than a
(residual) contribution to the masses of $r_1$, $i_1$, $c_1$ and $c_2$, the fermions, even at one-loop,
do not affect the inert-like minimum. The existence of the fermions thus introduces a qualitative
difference between both minima, and as such one must expect differences in the comparison of the depths
of the potential at each of them.

The minimisation procedure is like we already described, but now keeping all contributions --- scalar,
gauge and fermionic --- to the potential, its first derivatives and the one-loop masses. Once again we
require that the parameters are such that inert and inert-like minima coexist. The numerical scan of
the parameter space we considered tried to cover it as much as possible
--- namely, all quartic couplings were allowed to vary in their full range (constrained by bounded from below
and perturbative unitarity~\cite{Kanemura:1993hm,Akeroyd:2000wc, Swiezewska:2012yuk} constraints);
all masses at the inert minimum were allowed to take values according the the latest experimental
constraints.

The first difference one notices compared to the two previous cases is that it is far more difficult
to numerically solve the minimisation conditions for coexisting minima. This is presumably caused by the
``imbalance'' in the fermionic contributions, which contribute to one of the minima but not the other.
In fact, the values of the parameters which give an inert minimum may differ quite significantly from
those that give an inert-like one. Still, this observation is merely a technical one --- the parameter
 scan can be improved, yielding large number of valid points, by choosing appropriate initial numerical
 ``guesses'' in the numerical minimisation. Nonetheless, the added difficulty in finding coexisting minima
when fermionic contributions are included does point to this region of parameter space (where inert
and inert-like minima coexist) being more restricted than what occurred when only scalar or gauge contributions
were considered.

Another interesting observation concerns the fact that in the numeric scan we performed we found a greater
percentage of points for which the inert minimum lied above the inert-like one than what had been found
in the previous two cases. Again, this might simply be a case of a bias in our numerical sampling strategy.
But it is tempting to analyse the expressions for the one-loop potential, eqs.~\eqref{eq:V1}--\eqref{eq:Vt},
and try to verify whether this trend makes sense. In fact, the largest fermionic contribution to the one-loop
potential at the inert minimum is the top quark's, given by
\be
-\,\frac{3}{16\pi^2}\,  \,m_t^4(\mu) \left[\log\left(\frac{m^2_t(\mu)}{\mu^2}\right)\,-\,\frac{3}{2}\right]\,,
\ee
where $m_t(\mu)$ is the running mass, obtained from eq.~\eqref{eq:mtp}. With our choice $\mu$ = 200 GeV,
this gives a {\em large positive} contribution, of about $2.5\times10^7$ GeV$^4$. This value is large, when compared
with a zero contribution in the inert-like minimum, but not too large compared with the typical tree-level value of the potential, $\sim10^8$ GeV$^4$, so perturbation theory is not endangered. Thus we see that a
back-of-the-envelope calculation leads us to suspect that the value of the potential at the inert minima will have
positive contributions that the inert-like ones will be unaffected by.\footnote{Since physical predictions in
the one-loop potential formalism are guaranteed to be independent of the choice of renormalisation scale, this
is a sound calculation.}

\begin{figure}[ht]
\centering
\includegraphics[height=8cm,angle=0]{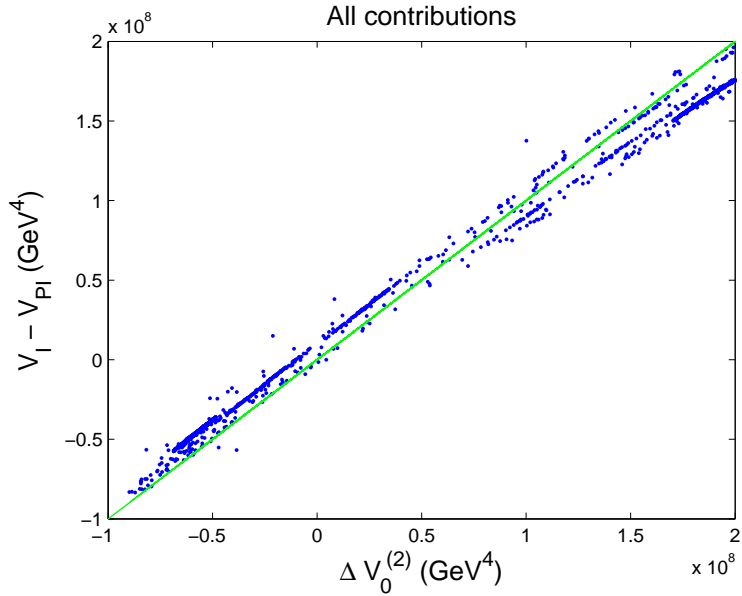}
\caption{One-loop computed difference in inert and inert-like minima depths ($V_I \,-\,V_{PI}$) {\em versus} the
tree-level expected depth difference given by eq.~\eqref{eq:difV02}, with scalar, gauge and fermionic contributions
taken into account.
}
\label{fig:dV2_F}
\end{figure}

Let us now consider the analog of fig.~\ref{fig:dV2_S}, where we compare the tree-level
expression for the relative depth of the potential at the inert and inert-like minima, $\Delta V_0^{(2)}$,
given by eq.~\eqref{eq:difV02}, but where now all masses, value of the minima and potential's
derivatives have incorporated, where appropriate, the fermionic contributions. The results are shown
in fig.~\ref{fig:dV2_F}. Clearly, the tree-level formula now constitutes a much cruder approximation
to the relative value of the minima of the potential than what occurred when only the scalar and gauge
contributions were included. We see that using $\Delta V_0^{(2)}$ one tends to underestimate the
difference in potential depths when the inert minimum is the global one, and overestimate it in
the opposite situation.

\begin{figure}[ht]
\centering
\includegraphics[width=.45\textwidth]{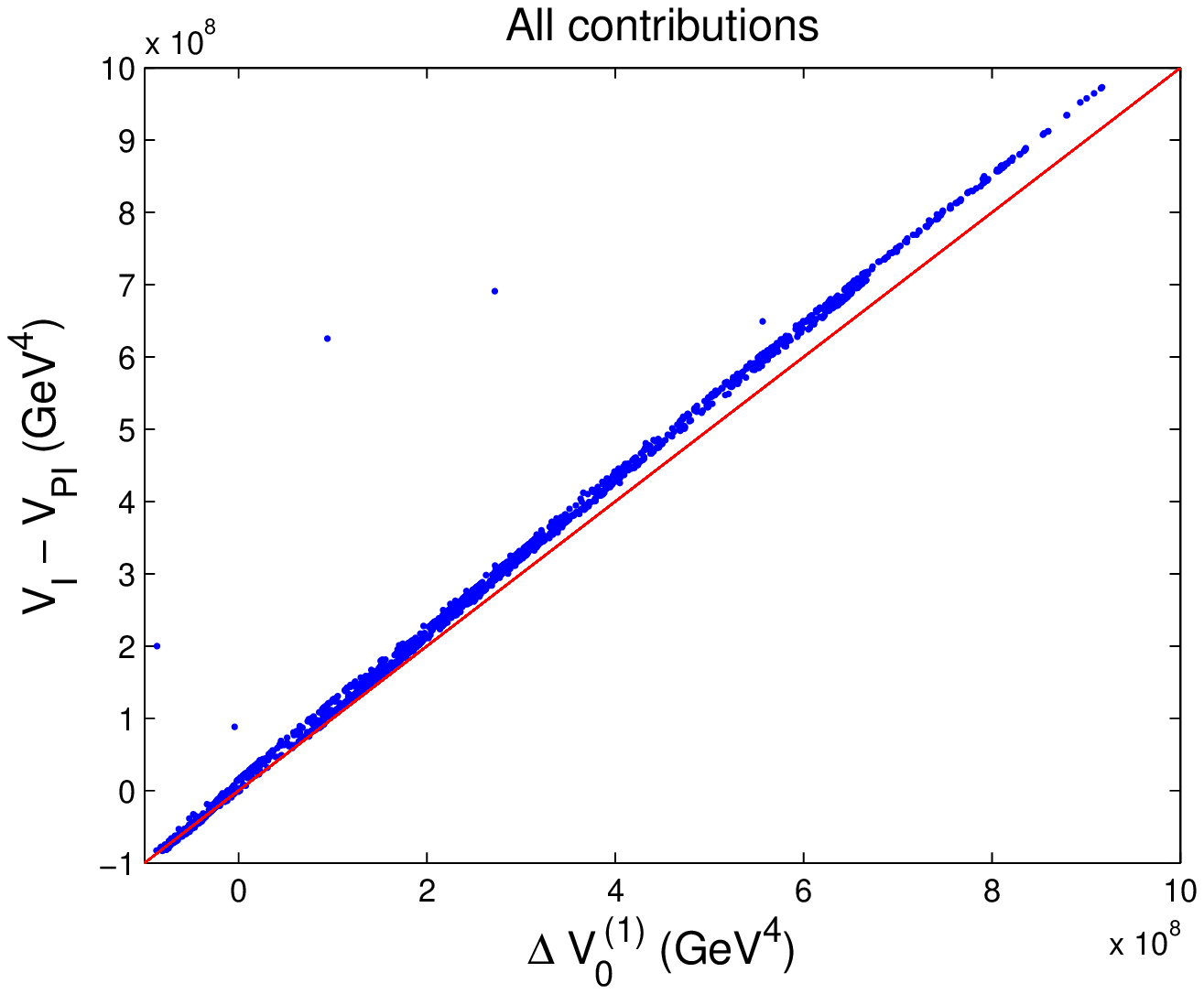}\hspace{.2cm}
\includegraphics[width=.45\textwidth]{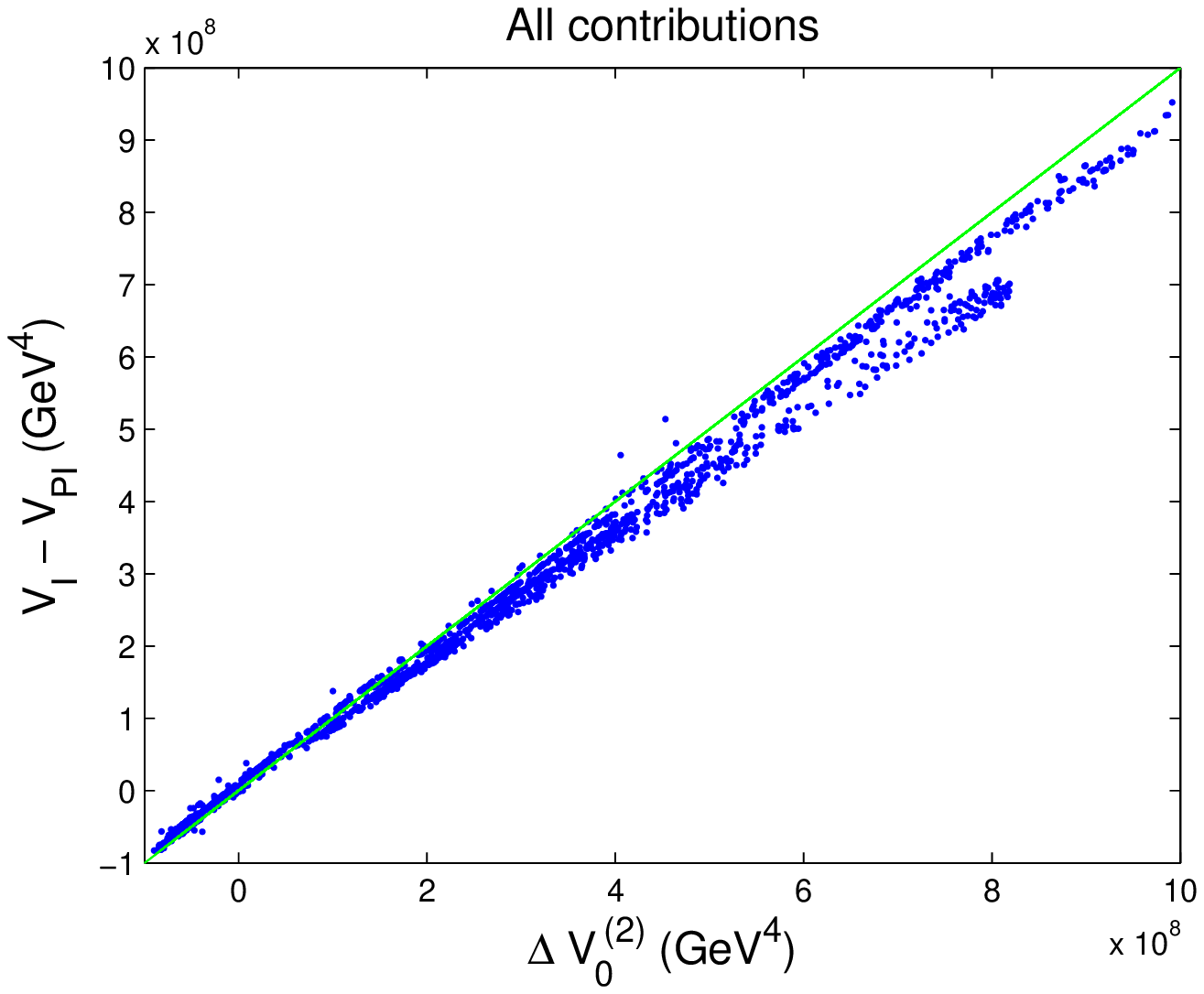}
\caption{One-loop computed difference in inert and inert-like minima depths ($V_I \,-\,V_{PI}$) {\em versus} the
tree-level expected depth differences given by eq.~\eqref{eq:difV01} (left) and eq.~\eqref{eq:difV02} (right),
with scalar, gauge and fermionic contributions taken into account. The straight lines in each plot
show what was to be expected if the tree-level results were exact.
}
\label{fig:compf}
\end{figure}

And again, the use of different tree-level expressions for the potential depths does not give
matching results.  In fig.~\ref{fig:compf} we show how $\Delta V_0^{(1)}$ and $\Delta V_0^{(2)}$
compare with the one-loop potential value differences. It is clear that $\Delta V_0^{(1)}$ tends to
underestimate the difference in depths of the potential for large values of this difference, whereas
$\Delta V_0^{(2)}$ tends to overestimate it.

\begin{figure}[ht]
\centering
\includegraphics[height=8cm,angle=0]{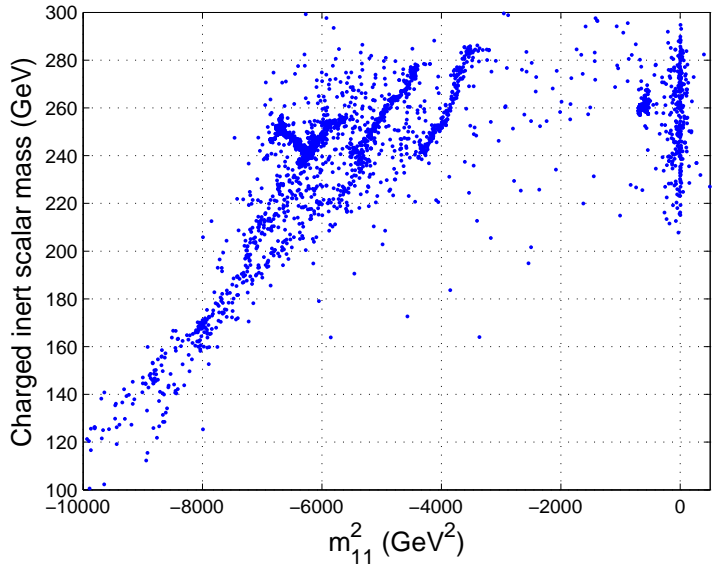}
\caption{The parameter $m^2_{11}$ {\em versus} the mass of the charged inert scalar at the
inert minimum. For all points shown, one-loop inert and inert-like minima coexist. At tree level, all
points with $m^2_{11} > 0$ would not be present.
}
\label{fig:m11}
\end{figure}

As before, the parameter space for which one-loop inert and inert-like minima coexist is larger than that
predicted by the tree-level analysis. For instance, consider the plot in fig.~\ref{fig:m11}. In it we plot the 
mass of the charged inert scalar ({\em i.e.} the charged scalar at the inert minimum) versus the value
of the $m^2_{11}$ parameter. We see that coexisting inert and inert-like minima occur even for $m^2_{11} > 0$
--- we had already seen, for $m^2_{22}$, in fig.~\ref{fig:m22_S}, this occurring when only the scalar 
contributions were included. Now we see that the parameter space for which coexisting minima may occur
is enlarged, compared to the tree-level expectations: in fact, coexisting minima were forbidden, at tree-level,
if $m^2_{11} > 0$, but the one-loop minimisation procedure shows that region is indeed 
allowed\footnote{Of course, coexisting minima with $m^2_{2} > 0$ are also possible, when one considers 
all contributions to the potential.}.
There is also a very small percentage of points for which the tree-level
predicted global minimum is inert (inert-like) and then found, at one-loop, to be inert-like (inert). However,
it is clear from the analysis of figs.~\ref{fig:dV2_F} and~\ref{fig:compf} that the tree-level formulae
are a much worse approximation to the one-loop results than in the cases where only the scalar and gauge sectors
were considered.

\section{Analysis of the results\label{sec:and}}

The thorough analysis of the past sections allowed us to compare the nature of the vacuum structure
of the IDM at tree level and one-loop level. We have shown that the tree-level formulae for the difference
in depths of the inert and inert-like minima yield different predictions than those obtained from
one-loop results. These differences are threefold, namely:
\begin{enumerate}
\item[(i)] Numerical, in the sense that the value of the relative depth of the minima is different at tree level
and one-loop. This, of course, is to be expected: we are working within a perturbation theory and thus
one-loop corrections to this quantity are normal.
\item[(ii)] Qualitative, in the sense that the global minimum of the potential for the IDM can be predicted
by the tree-level analysis to be inert/inert-like, and found, at one-loop, to indeed be inert-like/inert.
Thus the nature of the global minimum of the model can change, depending on the precision of the calculation.
\item[(iii)] Structural, in the sense that the parameter space allowed (for coexisting minima) at one-loop
is larger than that dictated by the tree-level analysis.
\end{enumerate}

The first of these conclusions, (i), might be relevant for eventual cosmological studies of the IDM. In fact,
one is not {\em forced} to require that the global minimum of the model be inert --- it could well happen that the
current vacuum wherein the universe inhabits is metastable, but with a tunnelling time to the deeper
unphysical inert-like minimum which is larger than the age of the universe. Conversely, there is also
the possibility that, during some time in the evolution of the universe since the Big Bang, the vacuum was,
for a brief period, inert-like, having then tunnelled very fast to a global inert minimum. Such tunnelling
times~\cite{Coleman:1977py,Rubakov,Adams:1993zs} depend very strongly on the difference in depth of the
potential at both minima, so an accurate knowledge of its value may be important.

The second conclusion is obviously more dramatic. As we have mentioned, it comes about only for
situations where
the inert and inert-like minima are close to degenerate (or more to the point, with a small depth difference
between them), at least if one bases one's tree-level predictions on formula~\eqref{eq:difV02}.
A tree-level analysis may lead us to believe that no tunnelling to a deeper inert-like minimum can occur, whereas the one-loop results would give the opposite prediction. This indicates that for such regions of parameter space the use of the one-loop analysis should
not be optional, but rather mandatory, since the tree-level results may be misleading.
We have also stressed that the several tree-level formulae for the relative depth of the minima
only agree at tree-level --- at one-loop the formulae are not equivalent, yielding slightly different results.
In particular, eq.~\eqref{eq:difV02} seems
to do a much better job than eq.~\eqref{eq:difV01}, as was mentioned concerning fig.~\ref{fig:dV1_S}: for
the former equation, there is some deviation from tree-level results in the one-loop results, but they remain
distributed around tree-level expectations; the latter equation, though providing a fair description of
one-loop results for most of the parameter space, does deviate strongly from them for a small
subset of points --- and those points are precisely those forbidden by the tree-level analysis, which brings
us to conclusion~(iii).

Some of these conclusions can be gleaned from the analysis made by the authors of ref.~\cite{Gil:2012}.
In fact, in their fig.~2 it can be seen that inert and inert-like minima can coexist even if
$\lambda_{345} < 0$, something which is forbidden at tree level. The analysis presented in
ref.~\cite{Gil:2012} did not undertake the study of inversions of the minima, though. In that study the
existence of an inert minimum is guaranteed by one of the renormalisation conditions, however it was not
enforced that the inert minimum is global at tree level. Thus it cannot be inferred from that analysis
whether or not the loop corrections change the nature of the global minimum of the potential.
It should also be stressed that the renormalisation
procedure used in~\cite{Gil:2012} was very different from the one we followed here. For instance, there
it was required that the one-loop inert minimum remains at the value $v_1^2 \,=\,
-2\,m^2_{11}/\lambda_1$, and other renormalisation conditions/inputs were used.
Direct comparison of our results
with theirs is therefore difficult, but certainly the results of~\cite{Gil:2012} confirm our finding (iii),
of a larger allowed parameter space at one-loop where inert and inert-like minima coexist.
Also, as in~\cite{Gil:2012}, we have discovered that the coexistence 
of  minima tends to occur in regions
of parameter space where the inert scalars (in the inert minimum) have relatively low masses, ranging
from roughly 300 to 400 GeV. It must however be emphasised that not all choices
of parameters which produce inert scalars with masses of this order lead to coexisting minima.

The conclusions of this study point to larger issues concerning the vacuum structure of the 2HDM. In fact,
expressions analogous to~\eqref{eq:difV01} and~\eqref{eq:difV02} were deduced~\cite{Ferreira:2004,Barroso:2005}
relating the depth of the tree-level potential at normal and charge breaking (CB) stationary points, or
at normal and CP breaking stationary points. It was then concluded that the existence of a normal minimum
precluded the existence of {\em any} CB or CP minima --- if such stationary points existed, they were guaranteed
to be saddle points. Conversely, if a CB or CP minimum existed, any normal stationary points would be saddle
points. The geometric analysis of ref.~\cite{Ivanov:2007} demonstrated that in fact minima which broke
different symmetries could not coexist in the 2HDM potential. However, general as that analysis was, it
relied entirely on the tree-level potential. The current work has shown beyond any doubt that the
conclusions derived at tree level for
a version of the 2HDM may not be wholly trusted once quantum corrections to the potential are taken
into account. This, then, begs the question: do the conclusions
of~\cite{Ferreira:2004,Barroso:2005,Ivanov:2007}
concerning the stability of the normal vacuum against tunnelling to CB or CP vacua stand at the one-loop
level? At the very least, the present work shows that one cannot be certain what the answer will be.

Unfortunately, trying to apply the effective potential methods to study CB/CP vacua coexisting with normal
minima will prove to be a complex challenge. One must not forget that, at tree level, the existence of a
normal minimum implies CP/CP stationary points to be saddle points (and vice-versa). Therefore one or more of
the squared scalar masses at those stationary points will perforce be negative, and we may anticipate
the one-loop effective potential having an imaginary component in the regions of parameter space one
would be interested in studying. There is no easy way out of this predicament --- the classic work of
 \cite{Fujimoto:1982tc} prescribes using the {\em convex envelope} of the potential in such cases, but
 it is not at all clear how one could proceed to construct such an entity when eight scalar fields are
 involved, as is the case of the 2HDM. Further, it could be argued that the prescription
 of~\cite{Fujimoto:1982tc} effectively precludes the existence of metastable minima, a possibility which
 should not be dismissed within the 2HDM. A possible way out of this conundrum is, once again, suggested
 by the results herein obtained: we have demonstrated that the parameter space allowed under one-loop
 minimisation conditions is larger than the tree-level allowed one. Thus, we may speculate on the following
 possibility: that, when using one-loop minimisation conditions to search for coexisting normal and CB/CP
 stationary points, they are actually solvable for values of the parameters not allowed under tree-level
 analysis --- and such parameters for which coexisting minima at one-loop therefore become a possibility. At the
 very least, the larger parameter space obtained for coexisting inert and inert-like minima at one-loop
 opens up this possibility.

Finally, a word on possible two-loop effects. The fact  that the inversion of inert and inert-like
minima was only found for regions of parameter space where the difference in depths of the potential at both
minima was small is reassuring, since it does not put into question the perturbative validity of the results.
In essence, what was found was a one-loop correction to a quantity that, at tree level, is comparatively small.
But that does raise the question of what the two-loop contributions might mean for these findings. Could
two two-loop effective potential undo the minima inversions we discovered at one-loop? The only way to
properly answer this question would be to perform a full two-loop calculation, and considering how hard to obtain
the current one-loop results were, such a calculation is clearly beyond the scope of the present work. We may,
however, argue that the two-loop contributions to the effective
potential~\cite{Martin:2001vx,Martin:2003it,Martin:2003qz} are expected to be much smaller than the one-loop ones.
Though they may introduce some numerical differences, one should not expect the two-loop contributions to alter
the fundamental qualitative differences we found using the one-loop potential: a larger region of parameter
space where minima can coexist; and the possible inversion of minima when quantum corrections to the potential
are added.

\section{Conclusions}
\label{sec:conc}

In short, the use of the one-loop effective potential to investigate the vacuum structure of the 2HDM
has been analysed for a specific version of the theory, the IDM. In order to do so the full one-loop
expressions for the scalars were computed and we have shown that the parameter
space for which inert and inert-like minima can coexist is larger than a simple tree-level analysis
would lead to believe. We have also shown that though the tree-level analysis provides a fair description
of the one-loop results for the difference in depths of the potential at both minima, care must be exercised
as to which tree-level formula one uses, as not all give the same answers. We have investigated the impact
of the scalar, gauge and fermionic sectors on the effective potential, and concluded that differences
between one-loop and tree-level results already emerge even if one considers only the contributions from
the scalar sector alone. The gauge sector produces effects which are qualitatively identical to the scalar one,
but the fermions do contribute quite differently to the inert and inert-like vacua and masses.
Further, for a small subset of
parameter space, the global minimum at one-loop is actually found to be of the opposite nature to what was
expected from the tree-level analysis. It therefore stands to reason that the one-loop effective potential
should be at the very least checked when considering issues of vacuum
stability within the 2HDM. These results cast doubts upon the validity, at higher orders, of the
tree-level conclusions regarding the stability of the 2HDM normal vacua against charge or CP breaking,
a question which remains open and which should be addressed.

\acknowledgments{
B{\'S} is very grateful to M.~Krawczyk for suggestions, fruitful discussions and reading the manuscript, and would also like to thank P.~Chankowski, G.~Gil, and M.~Sampaio for their help and discussions. This work was partially supported by the Polish National Science Centre grant PRELUDIUM under the~decision number DEC-2013/11/N/ST2/04214, and by the grant NCN OPUS 2012/05/B/ST2/03306 (2012-2016).
}

\appendix

\section{Special cases for $B$ functions}
\label{ap:specialB}

The $B$ functions presented in eqs.~\eqref{eq:B}, \eqref{eq:BSV} and~\eqref{eq:BVV} assume that
the masses circulating the loops of the scalars' self energies are different from zero. In fact
(specially in the case of the functions involving internal gauge boson lines, when photons
are involved) they seem singular in the limit of some of those masses going to zero, and/or in
the $p = 0$ regime. Those functions are however infrared safe, following the procedures explained by
Martin in~\cite{Martin:2003qz,Martin:2003it}. We present here some special cases which were helpful
during our calculations.

\paragraph{The $B$ function.}

The following formulae for $B$ (see eq.~\eqref{eq:B}) always assume, unless otherwise stated, $x\neq 0$,
$y\neq 0$, and are written in terms of the renormalisation scale $\mu$. For $p = 0$,

\begin{itemize}
\item $B(x,0,0)\,=\, \log\left(\frac{x^2}{\mu^2}\right)\,-\,1$.
\item $B(0,y,0)\,=\, \log\left(\frac{y^2}{\mu^2}\right)\,-\,1$.
\item $B(x,x,0)\,=\, \log\left(\frac{x^2}{\mu^2}\right)$.
\item $B(x,y,0)\,=\, -1\,+\,(1 + c)\log (1 + c) \,-\,c\log c\,+\,\log\left(\frac{a - b}{\mu^2}\right)$,
with $a = \mbox{max}(x^2,y^2)$, $b = \mbox{min}(x^2,y^2)$ and $c = b/(a - b)$.
\end{itemize}

For the $p\neq 0$ cases, let us define
\be
a\,=\,\frac{x^2}{p^2}\;\;\; , \;\;\; b\,=\,\frac{y^2}{p^2}\;\;\; , \;\;\; k_1 = a - b - 1
\;\;\; , \;\;\; k_2\,=\,b \,.
\ee
Then,
\be
x_{1,2} \,=\,\frac{1}{2}\,\left( - k_1 \,\pm\,\sqrt{k_1^2\,-\,4 k_2}\right)\;\;\; \mbox{with}
\;\;\; B_{x_i}\,=\,-1\,+\,(1 - x_i)\log (1 - x_i) + x_i \log (-x_i)
\ee
and we have
\be
B(x,y,p^2)\,=\,B_{x_1} \,+\,B_{x_2}\,+\,\log\left(\frac{p^2}{\mu^2}\right).
\ee
Notice this function is well-defined even in the limits $x_i\rightarrow 0$ or $x_i\rightarrow 1$.
Logarithms of negative numbers will of course give (expected) complex contributions to
the $B$ function.

\paragraph{The $B_{SV}$ function.}

The following formulae for $B_{SV}$ (see eq.~\eqref{eq:BSV}) always assume, unless otherwise stated,
$x\neq 0$, $y\neq 0$, and are written in terms of the renormalisation scale $\mu$. $x$ stands for
the scalar mass, $y$ for the gauge boson's.

For $p = 0$, in the Landau gauge, $B_{SV}(x,y,0)\,=\,0$. Otherwise:
\ba
B_{SV}(0,0,p) &=& -p^2\,\left[3\,B(0,0,p) + 2\right]\, ,
\nonumber \\
B_{SV}(x,0,p) &=& -3\,(x^2 + p^2)\,B(x,0,p)\,+\,3\,x^2\,\left[\log\left(\frac{x^2}{\mu^2}\right)
- 1\right]\,-\,2\,p^2\, ,
\nonumber \\
B_{SV}(0,y,p) &=& \frac{(y^2 + p^2)^2 \,-\,4\,p^2y^2}{y^2}\,B(0,y,p)\,-\,\frac{p^4}{y^2}\,B(0,0,p)
\,-\,(y^2 \,+\,p^2)\,\left[\log\left(\frac{y^2}{\mu^2}\right) - 1\right]\,.
\ea

\paragraph{The $B_{VV}$ function.}

The following formulae for $B_{VV}$ (see eq.~\eqref{eq:BVV}) always assume, unless otherwise stated,
$x\neq 0$, $y\neq 0$, and are written in terms of the renormalisation scale $\mu$.

We have, for the $p = 0$ case,
\ba
B_{VV}(x,y,0) &=&\frac{1}{4}\left[\log\left(\frac{x^2}{\mu^2}\right) - 1 \right]\,+\,
\frac{1}{4}\left[\log\left(\frac{y^2}{\mu^2}\right) - 1 \right]\,
\nonumber \\
 & & +\,\frac{1}{4x^2 y^2} \bigg\{ \left[(x^2 + y^2 - p^2)^2 +8 x^2 y^2\right]\,B(x,y,0) \,\nonumber\\
 &&-\,
 x^4\,B(x,0,0)\,-\,y^4\,B(0,y,0)\frac{}{}\bigg\} \\
B_{VV}(0,y,0) &=&\frac{3}{4}\left[\log\left(\frac{y^2}{\mu^2}\right) - 1 \right]\,+\,
\frac{9}{4}\,B(0,y,0)\,.
\ea
Otherwise the following special case is useful:\footnote{We only consider here
the case with one internal massless gauge boson, since there cannot be loops with
two internal photons.}
\be
B_{VV}(0,y,p) \,=\,\frac{3}{4}\left[\log\left(\frac{y^2}{\mu^2}\right) - 1 \right]\,+\,
\left(\frac{9}{4}\,-\,\frac{3 p^2}{4 y^2}\right)\,B(0,y,p)\,+\,\frac{3 p^2}{4 y^2}\,B(0,0,p)\,.
\ee


\bibliography{bibl}

\end{document}